\documentclass[10pt,aps,twocolumn,prl,superscriptaddress,noshowpacs,nofootinbib,noshowkeys,floatfix]{revtex4}
\usepackage[dvips]{graphics,graphicx}
\usepackage[colorlinks=true,linktocpage=true,linkcolor=blue,citecolor=blue,urlcolor=blue]{hyperref}
\usepackage[usenames,dvipsnames]{color}
\usepackage{amsmath, amssymb}
\usepackage{multirow}
\usepackage{bm}
\usepackage{color}
\usepackage[normalem]{ulem}  

\renewcommand\sout{\bgroup \color{blue} \ULdepth=-.5ex \ULset}
\usepackage{lipsum}
\usepackage{lipsum}

\usepackage{lipsum,graphicx,subcaption}
\newcommand{\comm}[1]{}
 
\begin{document}

\preprint{}

\title{
    Investigating the impact of extra resonance states in the van der Waals Hadron Resonance Gas Model

}      
\author{Nachiketa Sarkar}
\affiliation{School of Physical Sciences, National Institute of Science Education and Research, HBNI, Jatni-752050, India} 
 
\date{\today}

\begin{abstract}
 
  We investigate, in addition to the experimentally established hadrons, how the inclusion of extra resonance states, through the Hagedorn mass spectrum (HS) or Quark Model (QM) predicated states, affects the thermodynamic and transport quantities of the hadronic system in the van der Waals hadron resonance gas (VDWHRG) model. We found that the VDWHRG model with the HS provides the most accurate description of the lattice QCD results, both at zero and finite chemical potential. Moreover, the inclusion of these extra states has a significant impact on the van der Waals (VDW) parameters, which, in turn, affect the thermodynamic and transport quantities as well as the likely position of the liquid-gas phase transition critical point in the QCD phase diagram. Additionally, we infer that there is a strong correlation between the van der Waals parameters and the baryonic chemical potential. Overall, our study sheds light on the importance of considering extra resonance states and proper tuning of the VDW parameters in the VDWHRG model to enhance the accuracy and reliability of the model in the context of Ultra-relativistic heavy-ion physics.
 	
\end{abstract}

\maketitle

\section{I. Introduction}
The Lattice Quantum Chromodynamics (LQCD) \cite{Aoki:2006we,Borsanyi:2013bia,HotQCD:2014kol,HotQCD:2012fhj}, offers the most reliable equation of state for the partonic and hadronic phases of QCD matter at vanishingly small chemical potentials ($\mu_B$) and finite temperatures. According to LQCD simulation at low $\mu_B$, the phase transition from partonic to hadronic phase happens through a smooth  cross-over \cite{Aoki:2006we,Fukugita:1989yb,HotQCD:2014kol}. On the other hand, at high $\mu_B$, the first-order phase transition is predicted \cite{Gottlieb:1985ug,Herpay:2005yr,Pisarski:1983ms,Bowman:2008kc,Chatterjee:2012np,Bhattacharyya:2010wp}. Thus, there exist a critical point (CP) along the phase boundary at which the first-order phase transition ends and the cross-over begins. Theoretical and experimental efforts are ongoing to understand the phase boundary and find the location of the critical point \cite{Aoki:2006we,Aoki:2006br,Stephanov:2011pb,Stephanov:1998dy,Critelli:2017oub,Fodor:2004nz,Gavai:2004sd}. The Large Hadron Collider at CERN is dedicated on analyzing the properties of QCD matter produced at large temperatures and small baryonic chemical potentials. The Beam Energy Scan (BES) program of   Relativistic Heavy Ion Collider (RHIC) at Brookhaven National Laboratory (BNL) \cite{STAR:2009sxc}, tries to locate the QCD critical point by systematically reducing the collision energy. The BES-2 program \cite{Luo:2017faz}, has already begun which will reduce collision energy further and consequently reach higher chemical potential. Several other future experimental programs, including Compressed Baryonic Matter (CBM) \cite{Potel:2017upd} at the Facility for Antiproton and Ion Research (FAIR), the Nuclotron-based Ion Collider Facility (NICA) \cite{Sanchis-Alepuz:2015fcg}, and J-PARC-HI \cite{J-PARCHeavy-Ion:2016ikk}, will explore the properties of nuclear matter at high baryon density.

Thermal statistical model, like the Hadron Resonance Gas (HRG) model \cite{Karsch:2003zq, Bellwied:2013cta,Bellwied:2015lba}, is also very successful in reproducing the LQCD results of the hadronic phase of QCD matter at low temperatures ($<$ 150 MeV) and zero chemical potential. However, it fails to agree with LQCD results at higher temperatures, especially for higher-order fluctuations and correlations in the cross-over region \cite{Bazavov:2013dta,HotQCD:2018pds,Bellwied:2015lba,Alba:2017mqu}. Various modifications to the HRG model have been made to improve its agreement with LQCD results. One of the first extensions is the implementation of short-range repulsive interaction between the hadronic, popularly known as the excluded-volume (EVHRG) model \cite{Andronic:2012ut, Rischke:1991ke, Hagedorn:1980kb,Begun:2012rf, Yen:1997rv, Tiwari:2011km}. Other common approaches to incorporate interaction between the hadrons are the mean-field \cite{Huovinen:2017ogf,Pal:2020ucy,Steinert:2018zni}, and scattering phase shifts \cite{Venugopalan:1992hy,Steinert:2018zni,Friman:2015zua,Vovchenko:2017drx,Dash:2018can,Dash:2018mep}.
 
A recent extension of the EVHRG is the VDWHRG \cite{Vovchenko:2015xja}. In this formalism, both repulsive and attractive interactions are included by considering the van der Waals type of equation of state between the baryons. The grand canonical transformation and the quantum mechanical formulation of the full van der Waals equation are presented in \cite{Vovchenko:2015xja, Vovchenko:2015pya, Vovchenko:2015vxa}. The van der Waals parameters are fixed by reproducing the ground state properties of the nuclear matter. The value of the corresponding critical point of the nuclear matter is, $T = 19.7$ GeV and $\mu_B$ = 908 MeV \cite{Vovchenko:2015pya}. Finally, this formalism has been applied to the multi-component hadrons resonance gas using the same interaction constant \cite{Vovchenko:2016rkn}. Interestingly, this formalism describes the lattice results qualitatively better even in the cross-over region \cite{Vovchenko:2016rkn}.

Phase transition and criticality are built into the van der Waals equation, which exclusively depends on the choice of van der Waals constants. Following the reverse approach of \cite{Vovchenko:2016rkn}, the QCD critical point has been predicted, using the van der Waals constant extracted by comparing the lattice results at zero chemical potential \cite{Samanta:2017yhh}. Later, with a modified VDWHRG (repulsive interaction between mesons also considered) model and estimating the interaction constant based on finite lattice results, a discussion is made regarding the onset of the QCD phase transition and possible location of the critical point \cite{Sarkar:2018mbk}. It was argued that the change in degrees of freedom from the partonic to hadronic phase happened in the QCD phase transition but the HRG model only has the Hadronic phase. However, in the cross-over region, the VDWHRG model with tuned interaction parameters successfully describes the LQCD data which has both phases. So reinforced with lattice QCD, one can expect the VDWHRG model to mimic the onset of the QCD phase transition or the probable location of the critical point \cite{Samanta:2017yhh,Sarkar:2018mbk}.

Another major aspect of the HRG model is the number of resonance states considered in the calculation. Typically, all experimentally established discrete resonance states are taken into account in the analysis. However, previous studies have suggested that more resonance states, especially hadrons in the strange sector, are required for better agreement with lattice results \cite{Alba:2017mqu,Bollweg:2021vqf, Karthein:2021cmb}. Systematic studies have been performed in \cite{Alba:2017mqu, Karthein:2021cmb} by comparing the lattice results and HRG model calculation using different particle lists, namely: \\
 i) PDG16 (all well-established hadrons) \cite{ParticleDataGroup:2016lqr},\\
ii) PDG16+(Both well-and less established hadrons), \\
iii) QM (Quantum Model predicated hadronic list \cite{Capstick:1986ter,Ebert:2009ub,Alba:2017mqu}). \\
All of these studies indicate that additional states are required for a better description of the lattice results. To compensate experimentally yet confirmed heavy resonance states, another very popular approach is to consider the exponential mass spectrum proposed by Hagedorn \cite{Hagedorn:1965st,Hagedorn:1972spl,Hagedorn:1980kb,Hagedorn:1967tlw} along with the discrete resonance states. The inclusion of the Hagedorn states has a significant influence on the various thermodynamic quantities and transport coefficients of the hadronic system\cite{Noronha-Hostler:2014aia,Noronha-Hostler:2008kkf}. Previous studies have shown that simultaneous consideration of repulsive interactions and Hagedorn states improves LQCD-HRG agreement \cite{Vovchenko:2014pka,Sarkar:2017ijd}. Thus the inclusion of Hagedorn states or QM predicted states is expected to affect van der Waals constants and, consequently, other thermodynamic quantities or the location of the critical point. This motivated us to perform this study.\\

In this work, we consider two different versions of the HRG model:\\
a) VDW: HRG with van der Waals interaction between the baryons where as mesons are considered as a point particle \cite{Vovchenko:2016rkn}.\\
b) VDW+EV(Meson): Same VDW type of interaction for the baryon sector, but now mesons are considered as hard spheres with a radius of 0.2 fm\cite{Sarkar:2018mbk,Mohapatra:2019mcl}.\\
We also modify the particle list in two different ways:
I)  Inclusion of Hagedorn states along with established hadrons. \cite{Hagedorn:1972spl,Hagedorn:1980kb,Hagedorn:1967tlw}.\\
II). Use the QM-predicted hadronic states \cite{Bollweg:2021vqf}. \\
Finally, we study how the inclusion of the extra resonance states under these two different circumstances will influence the VDW parameters, consequently  various thermodynamic quantities at zero and finite chemical potentials. In this work, we use all well-established hadrons up to mass 3 GeV listed in   \cite{ParticleDataGroup:2016lqr} and the QM list is taken from \cite{Bollweg:2021vqf} . \\

The paper is structured as follows: Section II provides a brief description of the different models and observables used in this study. Section III presents our results and discussion. Finally, in Section IV, we summarize our findings for this work.

\section{I. Model  Description} 
 
\subsection{A. Ideal-HRG (IHRG) and Hagedorn States (HS)}  

 The simplest form of the HRG, known as ideal HRG (IHRG), is a multi-component, non-interacting thermally and chemically equilibrated statistical system. Considering the grand canonical ensemble of the hadronic gas, the logarithm of the partition function of the IHRG is \cite{Andronic:2012ut, Bhattacharyya:2013oya},
 \begin{equation}
 	\ln Z^{id}=\pm \sum_i 
 	\frac{Vg_i}{2\pi^2} \int_{0}^{\infty} p^2 dp \ln \Big[1 \pm \exp (-(E_i-\mu_i)/T)\Big].
 \end{equation} 
In this equation, each `i' represents the $i^{th}$ hadron, and the corresponding energy and chemical potential are denoted as $E_i=\sqrt{m_i^2+p^2}$ and $\mu_i=B_i\mu_B+S_i\mu_s+Q_i\mu_Q$, respectively. All other symbols have their usual meaning. The sign, `$\pm$', corresponds to the fermions and bosons, respectively. $\sum_i$, represents summation over all hadronic states. As mentioned previously, to compensate for missing high mass resonance states, one of our considerations is the continuous exponential mass spectrum as proposed by Hagedorn \cite{Hagedorn:1965st}. In this work, we used the following Hagedorn mass spectra \cite{Hagedorn:1972spl,Hagedorn:1980kb,Hagedorn:1967tlw},

\begin{equation}
 \rho(m)=C_H\frac{\theta(m-M_0)}{(m^2+m^2_0)^{a}}\exp{\frac{m}{T_H}}.
\end{equation} 
Where, `$m_{0} $' = 0.5 GeV, represents the lowest possible mass for a stable particle in the spectrum. $M_0 = 2$ GeV, is the minimum mass where Hagedorn contribution initiates. Parameter `$a$' is associated with the decay mode of the Hagedorn states. We have taken a = 5/4, which favors the multi-body decay of the heavier states\cite{Frautschi:1971ij}. Since we are comparing model results with those obtained from lattices up to 180 MeV temperatures, we fixed $T_H$ = 180 MeV in this analysis. $C_H$ is the only free parameter  \cite{Vovchenko:2014pka,Noronha-Hostler:2016ghw}. It should be noted that, in this analysis, we consider all the established discrete hadronic states (DH) with masses up to 3 GeV. However, when employing Hagedorn states, we take discrete hadrons with masses up to 2 GeV. Nevertheless, we have also checked by taking discrete mass up to 3 GeV along with Hagedorn states starting from 3 GeV ($M_0$ = 3 GeV). However, no significant differences were observed.

Including the Hagedorn states, the IHRG partition function has the following form \cite{Cleymans:2011fx, Kadam:2014cua},

\begin{equation}
	 \begin{split}
	\ln Z_{H}^{id}= \frac{V}{2\pi^2} \Bigg[ \pm  \sum_i g_i
	 \int_{0}^{\infty} p^2 dp \ln \big(1 \pm f_i \big) \\ + \sum_{\kappa  \in M,B,\bar{B}}  
	  \int_{M_0}^{\infty} \rho (m) dm \int_{0}^{\infty}  p^2 dp  \ln \big(1 \pm f_{\kappa} \big) \Bigg].
	  \end{split}
\end{equation} Here, M, B, and $\bar{B}$ represent meson, baryon, and anti-baryon, respectively, and $f = \exp\Big[-(E-\mu)/T\Big]$. Inline with the previous analysis \cite{Chatterjee:2009km,Kadam:2014cua}, in the present study also, we have used the same Hagedorn spectrum for meson, baryon, and anti-baryon. 
 Including the Hagedorn state, different thermodynamic quantities like pressure, number and energy density in the ideal HRG model reads \cite{Sarkar:2017bqy},

\begin{equation}
	\begin{split}
		P^{id}_H(T.\mu)= \frac{T}{2\pi^2} \Bigg[ \pm  \sum_i g_i
		\int_{0}^{\infty} p^2 dp \ln \big(1 \pm f_i \big) \\ + \sum_{\kappa  \in M,B,\bar{B}}   
		\int_{M_0}^{\infty} \rho (m) dm \int_{0}^{\infty}  p^2 dp  \ln \big(1 \pm f_{\kappa} \big) \Bigg],
	\end{split}
\end{equation}

\begin{equation}
	\begin{split}
		n^{id}_H(T,\mu)= \frac{1}{2\pi^2} \Bigg[ \pm  \sum_i g_i
		\int_{0}^{\infty} p^2 dp \frac{1}{f_i \pm 1}  \\ + \sum_{\kappa  \in M,B,\bar{B}}  
		\int_{M_0}^{\infty} \rho (m) dm \int_{0}^{\infty}  p^2 dp   \frac{1}{f_{\kappa} \pm 1}  \Bigg],
	\end{split}
\end{equation}

\begin{equation}
	\begin{split}
		\epsilon^{id}_H(T,\mu)= \frac{1}{2\pi^2} \Bigg[ \pm  \sum_i g_i
		\int_{0}^{\infty} p^2 dp \frac{E_i}{f_i \pm 1}  \\ + \sum_{\kappa  \in M,B,\bar{B}}  
		\int_{M_0}^{\infty} \rho (m) dm \int_{0}^{\infty}  p^2 dp   \frac{E_\kappa}{f_{\kappa} \pm 1}  \Bigg],
	\end{split}
\end{equation}

\subsection{B. Excluded-Volume HRG (EVHRG)}
 
 The excluded volume HRG (EVHRG) model incorporates repulsive interactions between hadrons by treating them as geometric objects with a hardcore radius, r. This approach results in modified expressions for the pressure and chemical potential, as shown in \cite{Rischke:1991ke}.
 
 \begin{equation}
 		P^{EV}(T,\mu_1,\mu_2,..)=  \sum_{i} P_i^{id} (T,\bar{\mu_1},\bar{\mu_2},..)
 \end{equation}

\begin{equation}
		 		\bar{\mu_i}=  \mu_1-V_{ev}P^{EV} (T,\mu_1,\mu_2,..)
		 		 \label{Eq:EVMU}
\end{equation}
 
Where, $V_{ev} =   \frac{16}{3} \pi r^3$ is the effective excluded volume for a hadron of radius r. Expressions for additional thermodynamic quantities can be derived using the EVHRG model as,

 \begin{equation}
	n^{EV}(T,\mu_1,\mu_2,..)=  \frac{\sum_i n^{id}_i (T,\bar{\mu_i})}{1+\sum_k V_{ev,k} n_k(T,\bar{\mu_k})}
\end{equation}

\begin{equation}
	\epsilon^{EV}(T,\mu_1,\mu_2,..)=  \frac{\sum_i \epsilon^{id}_i (T, \bar{\mu_i})}{1+\sum V_{ev,k} n_k(T,\bar{\mu_k})}
\end{equation}

\begin{equation}
	s^{EV}(T,\mu_1,\mu_2,..)=  \frac{\sum_i s^{id}_i (T,\bar{\mu_i})}{1+\sum_k V_{ev,k} n_k(T,\bar{\mu_k})}
\end{equation}

 \subsection{C. van der Waals HRG (VDWHRG)}

 The famous van der Waals equation is given by,
\begin{equation}
	p(T,n)= \frac{nT}{1-bn} - an^2,
\end{equation}
where, $n=(N/V)$ is the number density. The van der Waals constants a and b correspond to attractive and repulsive interactions respectively. Here also, $b=\frac{16}{3} \pi r^3$, is the excluded volume parameter for a hadron of radius r. The grand canonical formalism of the van der Waals equation of state of the hadron gas leads to the following modification of chemical potential \cite{Vovchenko:2015pya,Vovchenko:2015vxa,Vovchenko:2016rkn}
  \begin{equation}
\bar{\mu}=\mu - bp(T,\mu)- abn^2(T,\mu) + 2 an(T,\mu)
 \label{Eq:VDWMU}
\end{equation}

Where the pressure $p(T,\bar{\mu})$ and number density $n(T,\bar{\mu})$ related to each other by the following transcendental equation,
\begin{equation}
 p(T,\mu) = p^{id}(T,\bar{\mu}) - an^2(T,\bar{\mu})
\end{equation}

\begin{equation}
 n(T,\mu) = \frac{n^{id}(T,\bar{\mu})}{1+bn^{id}(T,\bar{\mu})}
\end{equation}

Other quantities like energy and entropy density can be obtained from well-known thermodynamic relations,

\begin{equation}
	\epsilon(T,\mu) =  Ts+\mu n -p
\end{equation} gives,
\begin{equation}
	\epsilon(T,\mu) = \frac{\epsilon^{id}(T,\bar{\mu})}{1+b\epsilon^{id}(T,\bar{\mu})} - an^2(T,\mu)
\end{equation}
 
\begin{equation}
	s(T,\mu) = \Big(\frac{\partial p}{\partial T}\Big)_\mu = \frac{s^{id}(T,\bar{\mu})}{1+bs^{id}(T,\bar{\mu})} 
\end{equation}

This formalism only takes into account the van der Waals interaction between fermions, while meson-meson, meson-baryon, or baryon-anti-baryon interaction is ignored. However, in ref. \cite{Sarkar:2018mbk,Mohapatra:2019mcl}, an EV type of interaction is considered for all mesons with a hardcore radius ($r_m$) of 0.2 fm. The partial pressure of the meson, baryon and anti-baryon components of a hadronic system in this formalism can be written as:

\begin{equation}
	P(T,\mu) =  P_{M}(T,\mu)+P_{B}(T,\mu)+P_{\bar{B}}(T,\mu)
\end{equation}

where 

\begin{equation}
	P_{M}(T,\mu)= 
	\begin{cases}
		$$ \mbox{\small $ \sum_{i \in M} P_i(T,\bar{\mu_i}^M)$}$$,  \text{\ if EV interaction consider }  \\
		$$ \mbox{\small $ \sum_{i \in M} P_i(T,\mu =0 )$ } $$ , \ \text{else} \\

	\end{cases}
\end{equation}

\begin{equation}
	P_{B}(T,\mu)=\sum_{i \in B} P_i(T,\bar{\mu_i}^B) 
\end{equation}

\begin{equation}
	P_{\bar{B}}(T,\mu)=\sum_{i \in B} P_i(T,\bar{\mu_i}^{\bar{B}}) 
\end{equation}

Where, modified baryons and anti-baryons' chemical potential read as (see Eq.\eqref{Eq:VDWMU}):
\begin{equation} 	
    {\mu_{(B,\bar{B})}^\star}(T,\mu)=\mu_{(B,\bar{B})} - bP_{(B,\bar{B})} - abn_{(B,\bar{B})}^2  + 2 an_{(B,\bar{B})}    
 	 \label{Eq:HRGVDWMU}.
 \end{equation}
  While the chemical potential for meson is modified according to Eq.\eqref{Eq:EVMU}. In this formalism, baryons and anti-baryons number density have the following form,

\begin{equation}
	n_{B,\bar{B}}=  \frac{\sum_{i \in {(B,\bar{B})}} n^{id}_i(T,\bar{\mu_i}^{B,\bar{B}})}{1+b\sum_{i \in {(B,\bar{B})}} n^{id}_i(T,\bar{\mu_i}^{B,\bar{B}})}
	 \label{Eq:HRGVDWN},
\end{equation}

\subsection{D. Observables : Specific Heat ($C_V$) Isothermal Compressibility ($\kappa_T$)  and Speed of Sound ($C^2_s$)  }

Thermodynamic observables, namely Specific heat ($C_V$), Isothermal comprehensibility ($\kappa_T$), or the Speed of Sound ($C^2_{s/n}$) at constant $s/n$, provide a wealth of information about the QCD phase transition or criticality. Specifically, the specific heat at constant volume, $C_V$, defined as $C_V=(\partial\epsilon/\partial T)_V$, is a crucial quantity that characterizes the equation of state of the system. The phenomenological significant of Cv is its association with temperature fluctuation which is further linked with the event-by-event fluctuation of mean $p_T$, thereby opening up the possibility of measuring the $C_V$ of the system formed in heavy ion collision.

Another phenomenologically relevant but less explored thermodynamic quantity in the contest of heavy ion collisions is the Isothermal-compressibility.  While Specific heat is related to temperature fluctuation, Isothermal-compressibility is associated with number fluctuation, and thus multiplicity fluctuation.  It's defined as the change in volume due to a change in pressure at a constant temperature, \cite{Mrowczynski:1997kz}

\begin{equation}
	\kappa_T |_{T,\langle N  \rangle}  = - \frac{1}{V} \Bigg(\frac{\partial V}{\partial P}\Bigg)_{T,\langle N  \rangle}.
\end{equation} 
 
 Where, $\langle N  \rangle$ is the average number of particle in the system.
 
 In the HRG model isothermal-compressibility can also be define as \cite{Mukherjee:2017elm},
 
 \begin{equation}
 	\kappa_T |_{T,\langle N  \rangle}  = \sum_{i (B,\bar{B}, M)}   \Bigg(\frac{\partial n_i}{\partial \mu_i}\Bigg)\Big/{n^2_i}  
 \end{equation} 

 Now, for the VDWHRG model from Eq.\eqref{Eq:HRGVDWMU} we get, 
  \begin{equation}
  	\Bigg(\frac{\partial {\mu_{(B,\bar{B})}^\star} }{\partial\mu_{(B,\bar{B})}}\Bigg)_T = \Bigg(1-bn_{(B,\bar{B}}  \Bigg)\Bigg[1+2a \Bigg(\frac{\partial  n_{(B,\bar{B})}  }{\partial\mu_{(B,\bar{B})}}\Bigg) \Bigg]
     \label{Eq:HRGMubyMuStar},
  \end{equation}

Using Eq.\eqref{Eq:HRGVDWN}, the partial  derivative  for the number density  $n (T, {\mu})$ with corresponding chemical potential at constant temperature can be written as,
 
\begin{equation}
		\begin{split}
	\sum_{i \in (B,\bar{B})} \Bigg(\frac{\partial {n_i} (T,\mu) }  {\partial\mu_i}\Bigg)_T =   \Bigg(1-bn_{(B,\bar{B})} \Bigg)^2 \\
		\Bigg(\frac{\partial {\mu_{(B,\bar{B})}^\star} }{\partial\mu_{(B,\bar{B})}}\Bigg)_T \sum_{i \in (B,\bar{B})}  \Bigg(\frac{\partial {n_i }(T,\mu^{\star}) }  {\partial\mu_i^{\star}}\Bigg)_T 
		\end{split}
	     \label{Eq:VDWHRGFluc},
\end{equation}
Now inserting Eq.\eqref {Eq:HRGMubyMuStar} in Eq.\eqref {Eq:VDWHRGFluc} and rearranging we get,

\begin{equation}
	\sum_{i \in (B,\bar{B})} \Bigg(\frac{\partial {n_i} (T,\mu) }  {\partial\mu_i}\Bigg)_T =  	\sum_{i \in (B,\bar{B})} \frac{\omega_i}{1-2a  \omega_i}.
	 \label{Eq:VDWHRGFluc2},
\end{equation}
Where, 
\begin{equation}
	 \omega_{i (B,\bar{B})} =  \Bigg(1-bn_{(B,\bar{B})} \Bigg)^3     \Bigg(\frac{\partial {n_{i (B,\bar{B})}} (T,\mu^{\star}) }  {\partial\mu_{i (B,\bar{B})}^{\star}}\Bigg)_T  
\end{equation}

In the meson sector we only consider EV interaction, i.e., a = 0. So for the meson, equation Eq.\eqref {Eq:VDWHRGFluc2}, (number density fluctuation) reduce to,

\begin{equation}
	\sum_{i \in M} \Bigg(\frac{\partial {n_i} (T,\mu) }  {\partial\mu_i}\Bigg)_T =  \omega_{M}
		 \label{Eq:VDWHRGFlucMeson},
\end{equation}
Of course, we will recover isothermal compressibility expression in the   IHRG model, if we put both a and b zero.
  
Finally using  Eq.\eqref {Eq:VDWHRGFluc2} and Eq.\eqref {Eq:VDWHRGFlucMeson}, the expression for the isothermal compressibility in the VDWHRG+EV model read as:

 \begin{equation}
	\kappa_T |_{T,\langle N  \rangle}  =\sum_{i \in (B,\bar{B})} \frac{\omega_i}{n_i^2(1-2a  \omega_i)} + \sum_{i \in M} \frac{\omega_i}{n_i^2} 
	 \label{Eq:ktFinal},
\end{equation} 

To calculate isothermal compressibility we use  Eq.\eqref {Eq:ktFinal} and for various HRG model we chose the a , b accordingly. \\

 
 
 

\begin{figure*}
	\includegraphics[height=3.8 in,width=7.4 in]{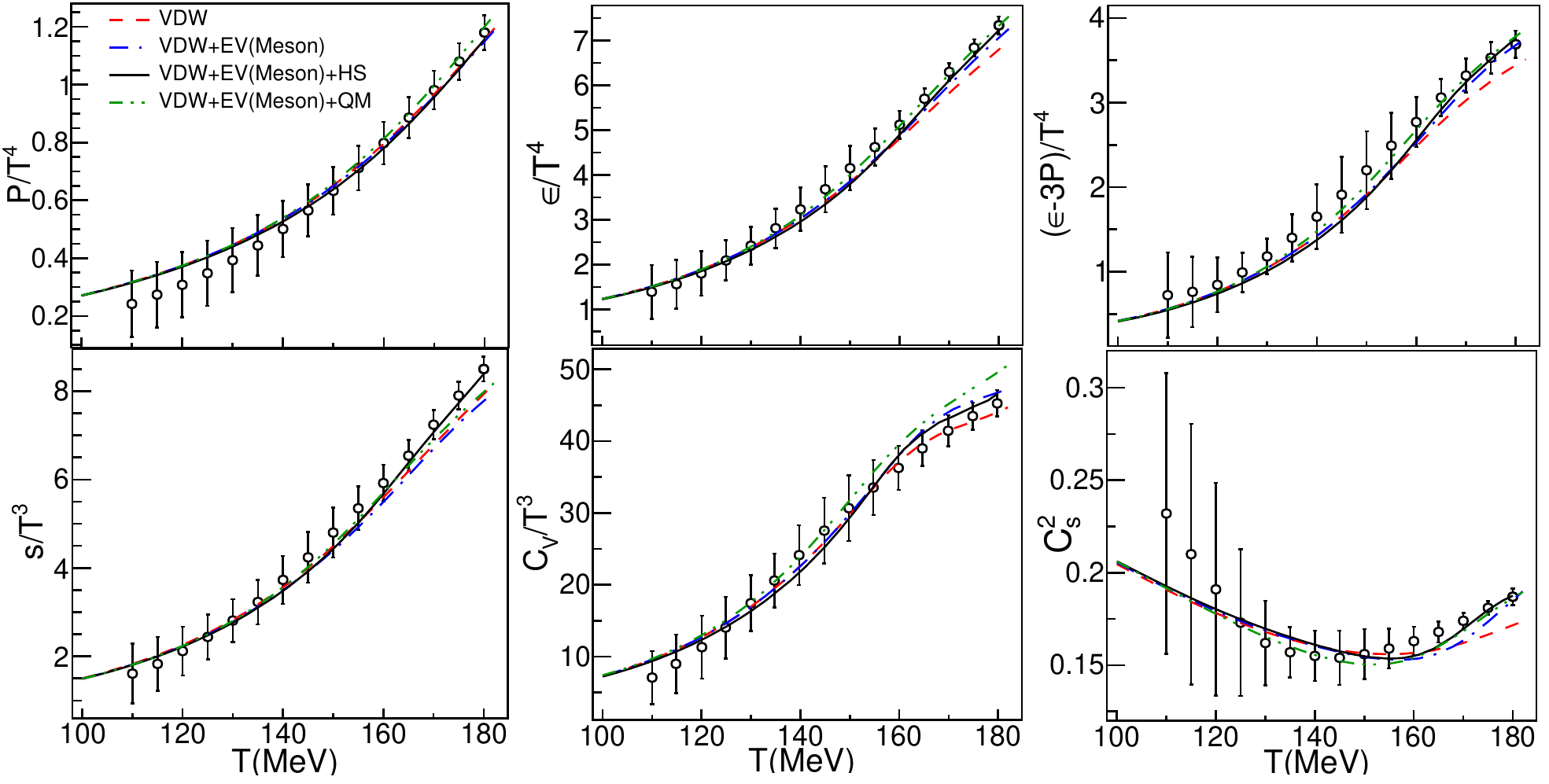}
	\caption{ A simultaneous fitting of the various thermodynamic quantities of the lattices results at zero chemical potential \cite{Borsanyi:2013bia} with different versions of the HRG model. Details of the various HRG models used in the fitting process are given in the text.}
	\label{fig:FitZeroMuB}
\end{figure*}

The speed of sound ($C^2_{s/n}$), which is defined as the derivative of pressure with respect to energy at constant entropy per particle,i.e., $C^2_{s/n} =  ({\partial p}/{\partial \epsilon})_{s/n}$,  is another thermodynamic quantity that is frequently discussed in heavy ion physics. Apart from providing the relevant equation of states for hydrodynamic simulation, it also serves as a sensitive indicator of phase transition or criticality \cite{Sorensen:2021zme}. The lattice calculations predict a weak dip around the cross-over temperature \cite{Borsanyi:2020fev, HotQCD:2018pds} which is considered an indication of QCD phase transition. The isentropic speed of sound can be computed using the formula, \cite{Floerchinger:2015efa}:
\begin{equation}
	C^2_{s/n} =  ({\partial p}/{\partial \epsilon})_{s/n}= \frac{n^2 \frac{\partial s}{\partial T}-2sn\frac{\partial n}{\partial T}+s^2\frac{\partial n}{\partial \mu}}{(\epsilon+p)\Big(\frac{\partial s}{\partial T}\frac{\partial n}{\partial \mu}-\big(\frac{\partial n}{\partial T }\big)^2\Big)}
	\label{Eq:Cs2}.
\end{equation}
Which reduce to $s/C_V$ at zero chemical potential.

\subsection{E. Transport Coefficient: Shear Viscosity ($\eta/s$)}

The transport coefficient that receives the most attention in heavy-ion physics is scaled shear viscosity, $\eta/s$. It relates to the momentum equilibration of the dissipative system and, therefore, to experimental observables like the elliptical flow coefficient. Understanding $\eta/s$ is crucial to gain insights into the QCD phase diagram. Similar to other thermodynamic quantities, $\eta/s$ exhibits a discontinuity during first-order phase transitions, whereas a smooth decline to a minimum value is observed at the crossover \cite{Csernai:2006zz, Lacey:2006bc} .

Although estimating the shear viscosity of a multi-component hadronic system is notoriously difficult, there have been numerous studies on the subject in the literature that rely on various approximations. These include the Chapman-Enskog (CE) \cite{Plumari:2012ep}  or relaxation time approximation (RTA)\cite{Gavin:1985ph,Gangopadhyaya:2016jrj} approach in relativistic kinetic theory, the Kubo formalism in linear response theory, microscopic transport calculations like UrQMD \cite{Demir:2008tr} and  SMASH \cite{Rose:2017bjz}, chiral perturbative theory (ChPT) \cite{Chen:2006iga}, and the quasi-particle approach \cite{Chakraborty:2010fr}. In our work, we adopted one of the most simplest approaches outlined in \cite{Gorenstein:2007mw}, to compute the $\eta/s$ of the hadron resonance gas system with an excluded volume correction. This method employs an analytically derived formula for estimating the shear viscosity of the hadronic gas mixture, which is given by,

\begin{equation}
	\begin{split}
		\eta^{vdW} =  \frac{5}{64\sqrt{8}} \sum_{i \in M,B,\bar{B}} \frac{<|P_i|> n^{vdW}_i(T,\mu)}{n^{vdW}(T,\mu), r^2_i}
	\end{split}
	\label{equ:DescritEta}
\end{equation}

Here , $P_i$ is the momentum  and $r_i$ is the hardcore radius of ith hadron respectively.

Several thermodynamic quantities have been found to be significantly influenced by HS states, Therefore, it is reasonable to assume that these states could also affect transport coefficients. Furthermore, because of their rapid decay properties, they are expected to influence the system's relaxation time hence the transport coefficients. However, our understanding of the quantum numbers associated with rapidly decaying massive Hagedorn states is limited, making it difficult to fully comprehend their contribution to shear viscosity. Despite this challenge, there have been efforts in the literature to calculate the transport coefficient for Hagedorn states \cite{Noronha-Hostler:2012ycm, Noronha-Hostler:2008kkf,Sarkar:2017bqy,Kadam:2014cua}. In ref.\cite{Noronha-Hostler:2008kkf}, the relaxation time was assumed to be inversely proportional to the decay width, and it was found that this resulted in a significant reduction in $\eta/s$ at the cross-over temperature.  Considering the Hadrons of hardcore radius r , shear viscosity Hagedorn states is given by \cite{Noronha-Hostler:2012ycm, Noronha-Hostler:2008kkf,Sarkar:2017bqy},

\begin{equation}
	\begin{split}
		\eta^{H}_{vdW} = \frac{1}{2 \pi^2 n_{vdW}^{H}}\frac{5} 
		{(64\sqrt{8}  r^2)} \\ \int{\rho(m) dm   \int_{0}^{\infty}  
			p^3dp   \Big(1\pm \exp\big(\frac{\epsilon-\mu}{T}\big)\Big
			)^{-1}} 
	\end{split}
	\label{equ:HSEta}
\end{equation}
 The total shear viscosity of the hadronic system consider both discrete and Hagedorn stats is,  

\begin{equation}
	\eta = \eta^{H}_{vdW}+\eta_{vdW}
\end{equation}

For simplicity in calculation, we disregard the contribution to the mean-free path resulting from interactions between discrete states and Hagedorn states due to the much lower population of HS compared to discrete states.


 \begin{table}[t]
 	\centering
 	\begin{tabular}[t]{clcc}
 		\hline
 		& \multicolumn{2}{c}{Zero Chemical Potential }&\\

 		\hline
 		VDW-a&VDW-b&   $C_{H}$&$\chi^2$/ndf\\
 		\hline
 		
 		\multicolumn{4}{c}{VDW}\\
 		\hline
 		\hline
 		1.25 $\pm$  0.015&5.78 $\pm$ 0.08 &- -&0.65\\
 		
 		\hline
 		\multicolumn{4}{c}{VDW+EV ($r_m$ = 0.2 fm)}\\
 		\hline
 		\hline 
 		1.02 $\pm$  0.015&	4.46 $\pm$ 0.05 &- --& 0.28 \\
 		
 		\hline
 		\multicolumn{4}{c}{VDW+EV ($r_m$ = 0.2 fm)+HS}\\
 		\hline
 		\hline
 		1.33  $\pm$ 0.018  & 5.30 $\pm$ 0.07  & 0.060 $\pm$ 0.003 & 0.15\\
 		\hline
 		\hline
 		\multicolumn{4}{c}{VDW+EV ($r_m$ = 0.2 fm)+QM}\\
 		\hline
 		\hline
 		0.861  $\pm$ 0.016 &5.56 $\pm$ 0.07  & --   -- & 0.42\\
 		\hline
 		
 	\end{tabular}
 	
 	\caption{Simultaneous fit results of lattice QCD data at zero baryon chemical potential using different variants of the  HRG model. }
 	\label{Tab:FiTSumZeroMuB}
 \end{table}

 \begin{table}[t]
 	\centering
 	\begin{tabular}[t]{clcc}
 		\hline
 		& \multicolumn{2}{c}{Finite Chemical Potential }&\\

 		\hline
 		VDW-a&VDW-b&$C_{H}$&$\chi^2/ndf$\\
 		\hline
 		
 		\multicolumn{4}{c}{VDW}\\
 		\hline
 		\hline
 		0.842 $\pm$  0.020&3.80 $\pm$ 0.05 &--&0.30\\
 		
 		\hline
 		\multicolumn{4}{c}{VDW+EV ($r_m$ = 0.2 fm)}\\
 		\hline
 		\hline
 		0.735 $\pm$  0.011&3.08 $\pm$ 0.04 &---&  0.39\\
 		
 		\hline
 		\multicolumn{4}{c}{VDW+EV ($r_m$ = 0.2 fm)+HS}\\
 		\hline
 		\hline
 		0.710  $\pm$ 0.012  	&3.58 $\pm$ 0.05  & 0.065 $\pm$ 0.002 &0.11\\
 		
 		\hline
 		\hline
 		\multicolumn{4}{c}{VDW+EV ($r_m$ = 0.2 fm)+QM}\\
 		\hline
 		\hline
 		0.702  $\pm$ 0.015  	&3.33 $\pm$ 0.06  & -- & 0.37\\
 		\hline
 	\end{tabular}
 	
 	\caption{Same as the caption of the Tab.\eqref{Tab:FiTSumZeroMuB} but for the finite chemical potential }
 	\label{Tab:FitFiniteMuB}
 \end{table}

 \section{III. Result and Discussion}
 

We extract the van der Waals parameters by simultaneously fitting the zero and finite chemical potential lattice results of various thermodynamic quantities for different versions of HRG models: I) VDW, II) VDW+EV(Meson), III) VDW+EV(Meson)+HS, and IV) VDW+EV(Meson)+QM. To get the best fit of the model to the data, we use the $\chi^2$ minimization technique, which is defined as follows:

\begin{equation}
	\chi^2= \frac{1}{N} \sum_i \sum_j \frac{(R^{LQCD}_{i,j}-R_{i,j}^{Model})^2}{(\delta R^{LQCD}_{i,j})^2}
\end{equation}

 In this equation, the variables $R^{LQCD}{i,j}$ and $R_{i,j}^{Model}$  correspond to the $i^{th}$ data point of the $j^{th}$ observable for the  LQCD simulation and the model result, respectively. The symbol $\delta R^{LQCD}_{i,j}$ represents the error associated with the LQCD result for the respective data point. $N$ is the total number of LQCD data used in the fitting process.
 \begin{figure*}[t]
 	\centering
 	\includegraphics[width=0.8\linewidth]{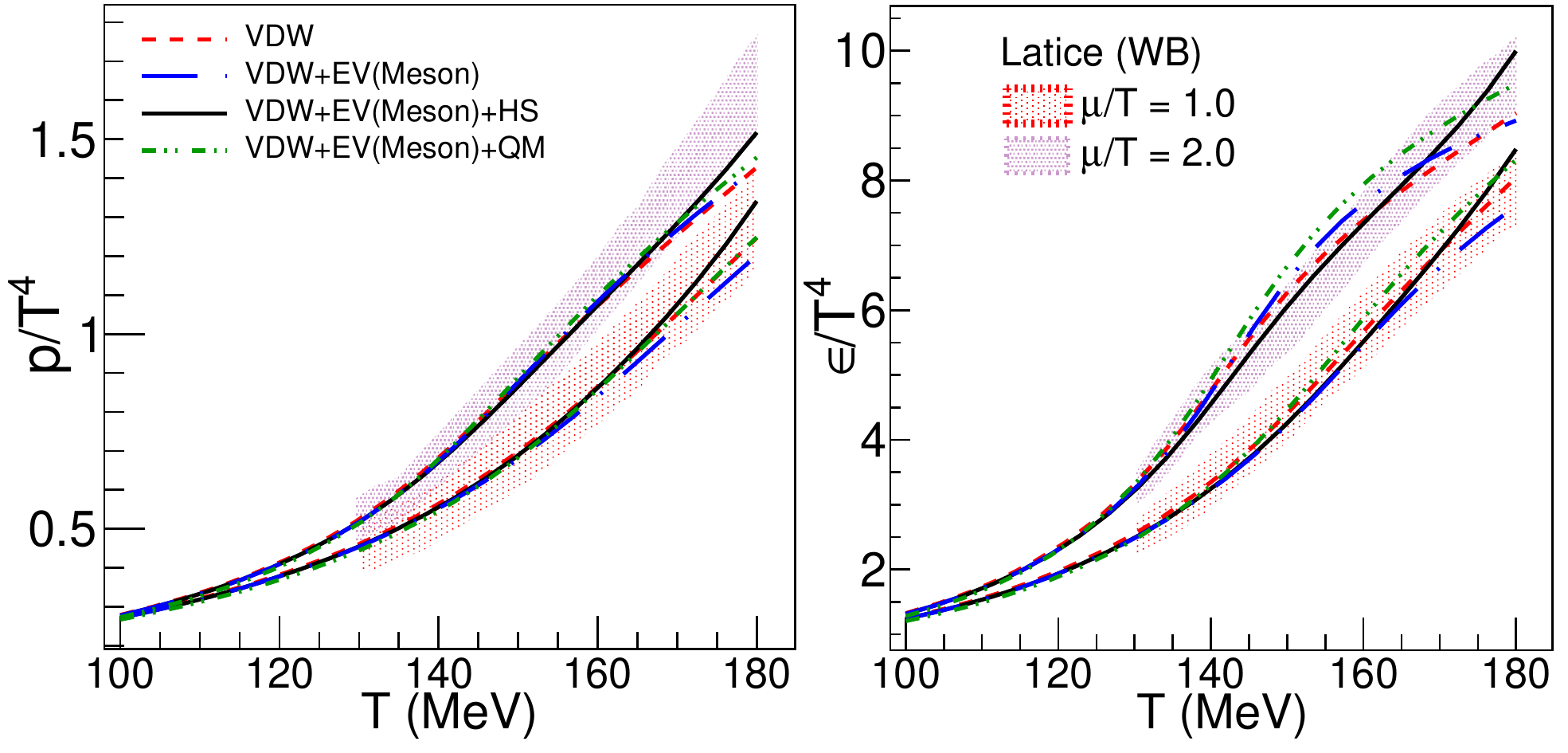}
 	\caption{Same as the caption of the Fig\eqref{fig:FitZeroMuB}, but with the finite $\mu_B$ lattice results \cite{Bazavov:2017dus}.}
 	\label{fig:FitFiniteMuB}
 \end{figure*}

\subsection{A. Fitting At Zero $\mu_B$} 
At zero chemical potential, we use the lattice results of various thermodynamic quantities such as $p/T^4, \epsilon/T^4, (\epsilon -3p)/T^4, s/T^3, C_v/T^3$, and $C^2_s$, obtained from \cite{Borsanyi:2013bia}, in the fitting process. To cover the crossover region \cite{HotQCD:2012fhj,Bazavov:2011nk, Borsanyi:2010bp}, we chose a temperature range for the lattice data from 110 to 180 MeV. This part of the analysis is presented in figure \eqref{fig:FitZeroMuB}, and the fitting summary is tabulated in Tab.\eqref{Tab:FiTSumZeroMuB}.

As can be seen from figure \eqref{fig:FitZeroMuB} or the $\chi^2$ value from the Tab.\eqref{Tab:FiTSumZeroMuB}, the VDWHRG model with Hagedorn states (VDW+EV(Meson)+HS) describes the lattice results of all the thermodynamics quantities best, compared to others. Many studies have been conducted using different $T_H$ values \cite{Majumder:2010ik, Noronha-Hostler:2012ycm, Vovchenko:2014pka, Lo:2015cca}, resulting in different Hagedorn parameters. As already mentioned, in this study, we use $T_H$ = 180 MeV, and the corresponding fitted value of the parameter, $C_H$ is $\sim$ 0.06 $\text{GeV}^{3/2}$. In a previous study using the same $T_H$, the value of $C_H$ is $\sim$ 0.63 $\text{GeV}^{3/2}$ \cite{Majumder:2010ik}, which is quite large compared to our analysis. The reduction of the $C_H$ parameter can be attributed to the incorporation of attractive interaction in the HRG model in the present study. It is worth noting that attractive interactions and Hagedorn states have a qualitatively similar effect on the thermodynamic quantities, and in this sense, attractive interactions can be viewed as mimicking the effect of additional states.
  
As expected, the Hagedorn states, with their high mass, have a significant impact on energy, entropy, and $C^2_{s}$, especially around the cross-over temperature, while having a minimal effect on pressure \cite{Vovchenko:2014pka}. The VDWHRG model with QM states (VDW+EV(Meson)+QM) also provides a good description of the lattice results. Furthermore, due to the presence of extra low mass states, it has a noteworthy impact on thermodynamic quantities, even at low temperatures. As evident in terms of the $\chi^2/ndf$ \eqref{Tab:FiTSumZeroMuB}, the VDW+EV(Meson) shows better agreement with the lattice results than the VDW model. This underscores the importance of meson interactions in the vanishingly small $\mu_B$ region, where mesons dominate. All four different VDWHRG models considered, exhibit a dip in $C^2_{s}$ at approximately the same temperature. However, the VDWHRG model with HS and QM predicated states describes the lattice data more accurately beyond the cross-over point. It can be easily followed from the definition of $C^2_{s} = s/C_V$. At high temperature, due to the availability of heavier resonance states in both the VDWHRG model with QM or HS states, the energy of the system is mostly used to create heavier resonance states, resulting in a more rapid increase in entropy compared to $C_V$. Therefore, $C^2_{s}$ increases after the crossover temperature of $\sim 156$ MeV, making the model calculation of $C^2_{s}$ in the VDWHRG framework with HS and QM states closer to the lattice result. As shown in Tab.\eqref{Tab:FiTSumZeroMuB}, fitted value of the repulsive parameter in the VDW model is $b=5.76 \ \text{fm}^3$, corresponding to a hardcore radius of $r=0.7 \ \text{fm}$, and the attractive parameter value is $a=1.25\ \text{MeV.fm}^3$, which is in agreement with the value reported in \cite{Samanta:2017yhh}. In the VDW+EV(Meson) model, the mesons also contribute to the repulsive interaction, resulting in a reduction of the effective hardcore radius of the baryons to 4.46 fm. However, including QM or HS states leads to an increase in the number density, which is counteracted by an increase in the $b$ value in the corresponding models.  Off course the dependence of the van der Waals parameters on the number of states are non-trivial.

\subsection{B. Fitting At Finite $\mu_B$} 

We now turn our attention to the finite chemical potential region of the QCD phase diagram with the objective of fitting the lattice results for pressure and energy density at finite $\mu_B$. Specifically, we use the lattice results \cite{Bazavov:2017dus} for two different cases, $\mu_B/T=1$ and $\mu_B/T=2$, up to a temperature of 180 MeV, covering the $\mu_B$ range from 100 to 360 MeV. The comparison of the model calculations with the lattice results is shown in figure \ref{fig:FitFiniteMuB}, and the analysis findings are summarized in Tab. \ref{Tab:FitFiniteMuB}. The VDW+EV(Meson)+HS model best describes the lattice results, as it does at zero $\mu_B$. Notably, the van der Waals parameters exhibit a significant decrease, indicating a strong dependence on chemical potential. Unlike at zero $\mu_B$, the VDW model fits the lattice data better than the VDW+EV(Meson) model at finite $\mu_B$, implying that meson interactions may not be significant in the baryon-dominated region. The $\chi^2$ value for the VDW+EV(Meson)+QM model has slightly worsened compared to the VDW model, as it overestimates the energy density at low-temperature regions. However, all variants of the VDWHRG model underestimate the pressure at high $\mu_B$.

 \begin{figure}[]
	
	\includegraphics[height=5.2 in,width=3.in]{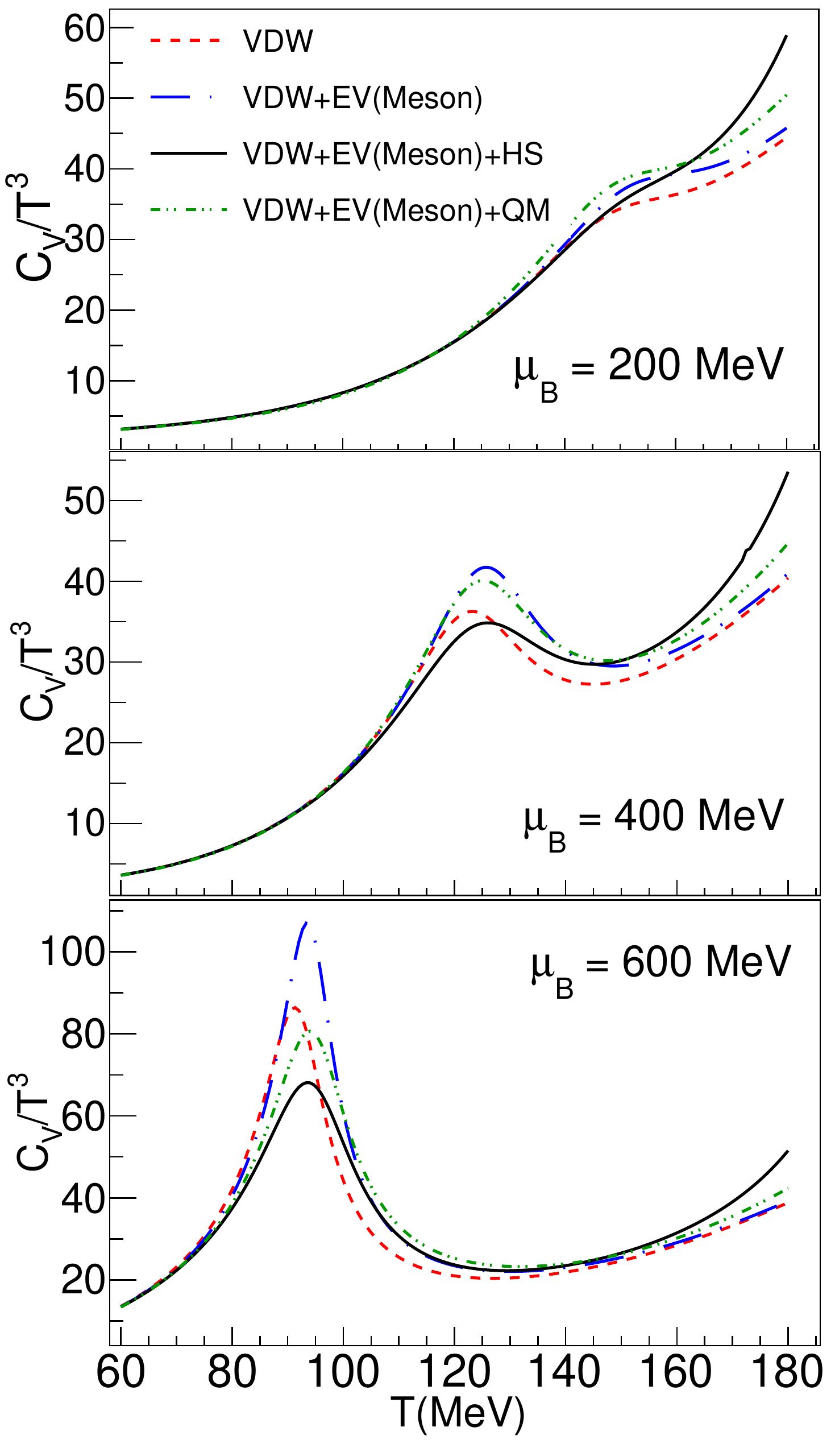}

	\caption{ The scale-specific heat as a function of temperature is measured in different versions of the HRG model at three distinct chemical potentials, as specified in the legend. The VDW parameters for the corresponding VDWHRG models are given in \eqref{Tab:FitFiniteMuB}.  }
	\label{Fig:Cv1}
\end{figure}

\begin{figure}[]
	\includegraphics[height=5.2 in,width=3.in]{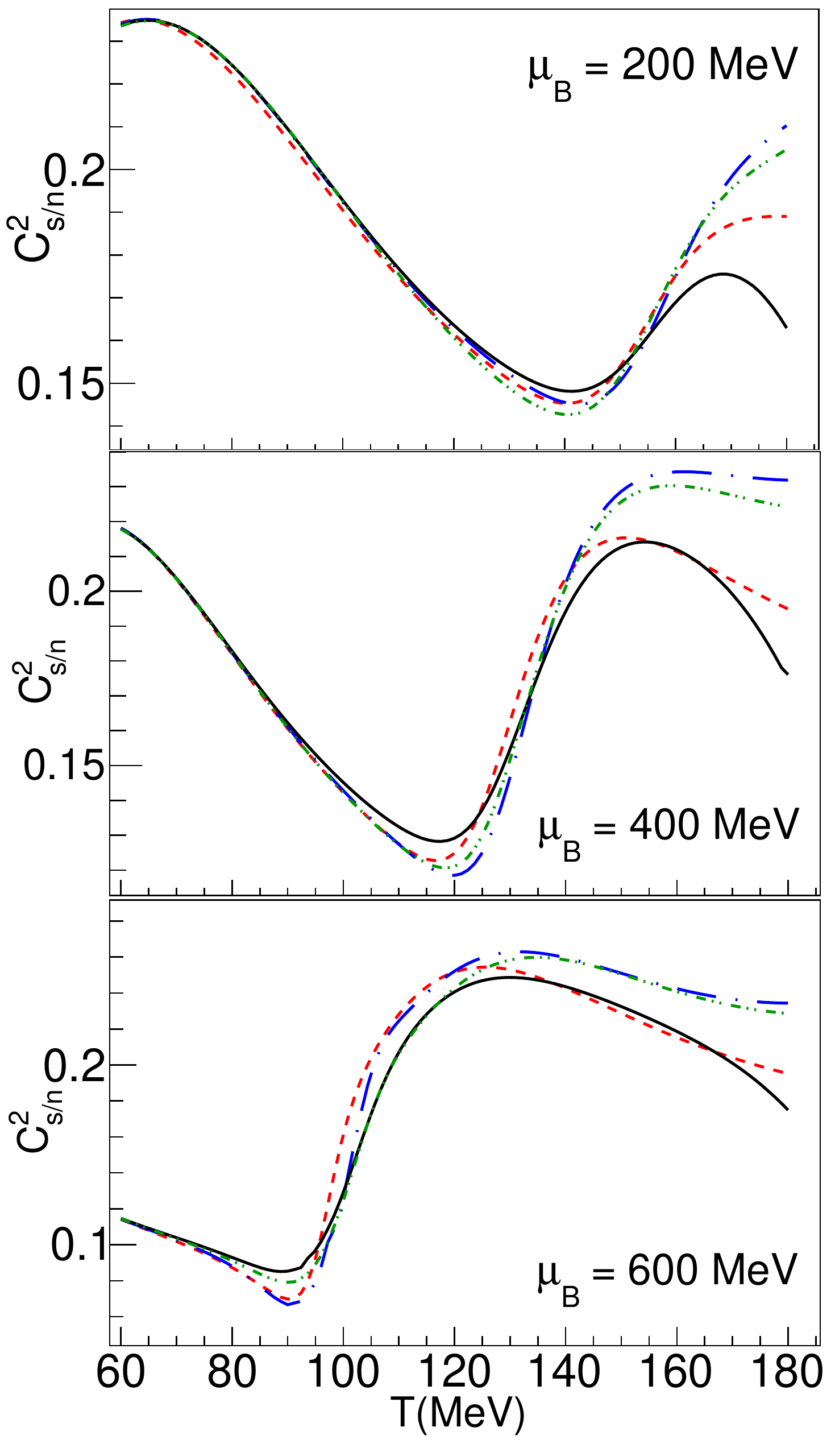}

	\caption{The speed of sound at constant $s/n$ as a function of temperature for various versions of the HRG model, measured at three different chemical potentials as indicated in the legend. The VDW parameters for the corresponding VDWHRG models are given in \eqref{Tab:FitFiniteMuB}. }
	\label{Fig:Cs1}
\end{figure}

 \begin{figure*}[t]
	
	\includegraphics[height=2.2 in,width=5.4in]{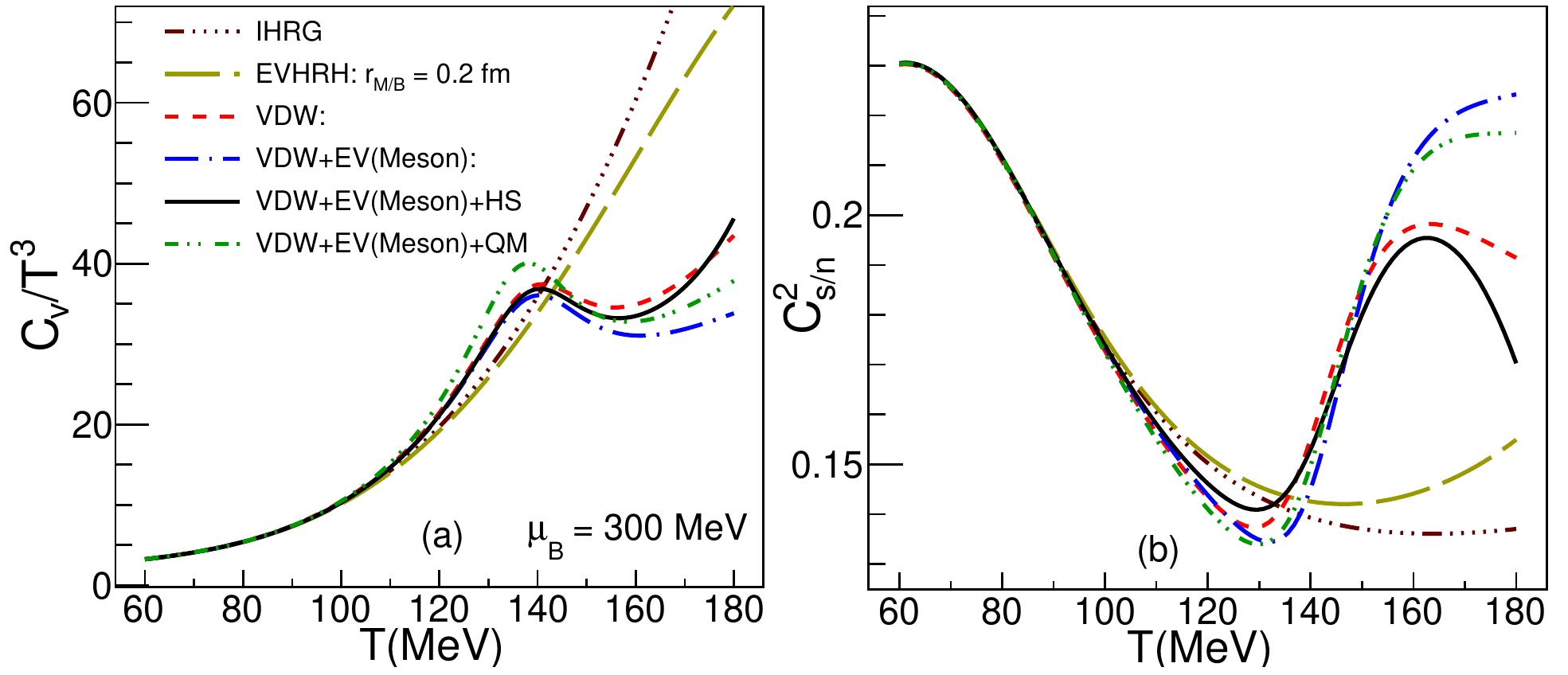}

	\caption{Temperature dependence of a) scale specific heat, and  b) $C^2_{s/n}$, at $\mu_B$ = 300 GeV, for different HRG models as specified in the legend. The same VDW constants were used for all the VDWHRG models, and the results of IHRG and EVHRG were also included for comparison.}
	\label{Fig:CsandCvVsT}
\end{figure*} 

\subsection*{C. Specific Heat And Speed Of Sound }

After fixing the van der Waals parameters for both zero and finite chemical potential, our subsequent interest is to investigate how the parameters tuned for different HRG models under different conditions affect various thermodynamic observables and transport coefficient.

Figures \eqref{Fig:Cv1} and \eqref{Fig:Cs1}, respectively illustrate the variation of the scaled specific heat $C_V/T^{3}$ and the speed of sound $C_s^{2}$ as a function of temperature for different versions of the VDWHRG model used in this study at three different chemical potentials, $\mu_B = $ 200, 400, 600 MeV.

In the low-temperature regime, with an increase in temperature, new resonance states emerge, leading to a rise in the system's energy. Consequently, $C_V/T^{3}$ steadily increases, while the speed of sound decreases. As the system approaches phase transition temperature, system energy rapidly increases,  causing, $C_V$ to either diverge or exhibit a peak, while $C_s^{2}$ reaches a minimum value, commonly referred to as the soften-point in the equation states which act as an indicator of  
QCD phase transition \cite{Hung:1994eq}. The appearance of a peak in $C_V$, based on Gibbs' order of phase transition criteria, indicates a second-order phase transition. As expected, the position of the peak or the minimum in $C_s^{2}$  shifts to the lower temperature region as we increase the chemical potential. Following the phase transition, the temperature of the system rises rapidly, causing an increase in pressure that dominates over the increase in system energy, resulting in a rise in $C_s^{2}$  and a decrease in $C_V$.  At extremely high temperatures, the production of heavy resonances causes another increase in $C_V$ and a decrease in $C_s^{2}$. This effect is most notable in the VDWHRG model with Hagedorn states. The effect of van der Waals parameters is more pronounced at higher chemical potentials.
As shown in the figure \eqref{Fig:Cv1} and \eqref{Fig:Cs1} , due to the lower value of the repulsive interaction constant, the VDW+EV model exhibits the highest peak value in $C_V$ or the lowest minimum at the soften-point for $C_s^{2}$ at $\mu_B$ = 600 and 400 MeV. On the other hand, the VDWHRG model with QM predicted hadronic states dominates in the intermediate temperature and low $\mu_B$ = 200 MeV regions. The phase transition point (the peak) varies slightly among the various HRG models considered here owing to the different van der Waals parameters taken into account. 

The differences in the temperature profile of $C_V$ or $C_s^{2}$ among these HRG models are the results of the combined impact of different hadronic states or VDW constants. To illustrate this point further, we plotted $C_V$ and $C_s^{2}$ as functions of temperature for all four VDWHRG model variants, using the same VDW constant, in figure \eqref{Fig:CsandCvVsT}-(a,b), respectively. There were no significant differences observed in $C_V$ or $C_s^{2}$ among the different HRG models at low temperatures. As expected, the peak position in $C_V$ and the location of the softened point in $C_s^{2}$ are almost identical for all VDWHRG models. However, because of the additional low-mass states in the QM model, the VDW+EV(Meson)+QM model has the highest peak in $C_V$ or the lowest minimum in $C_s^{2}$. At high temperatures, the repulsive interaction between mesons leads to a significant reduction in $C_v$ in the VDW+EV(Meson) model compared to the VDW model. While the additional states introduced by HS or QM have a prominent counter-effect. For completeness,  we also included IHRG and EVHRG results in figure \eqref{Fig:CsandCvVsT} to compare those with VDWHRG model results. As expected, no such peak appears in either of the models.

\subsection{D. Isothermal Compressibility}
\begin{figure*} 
	\includegraphics[height=2.2 in,width=5.4in]{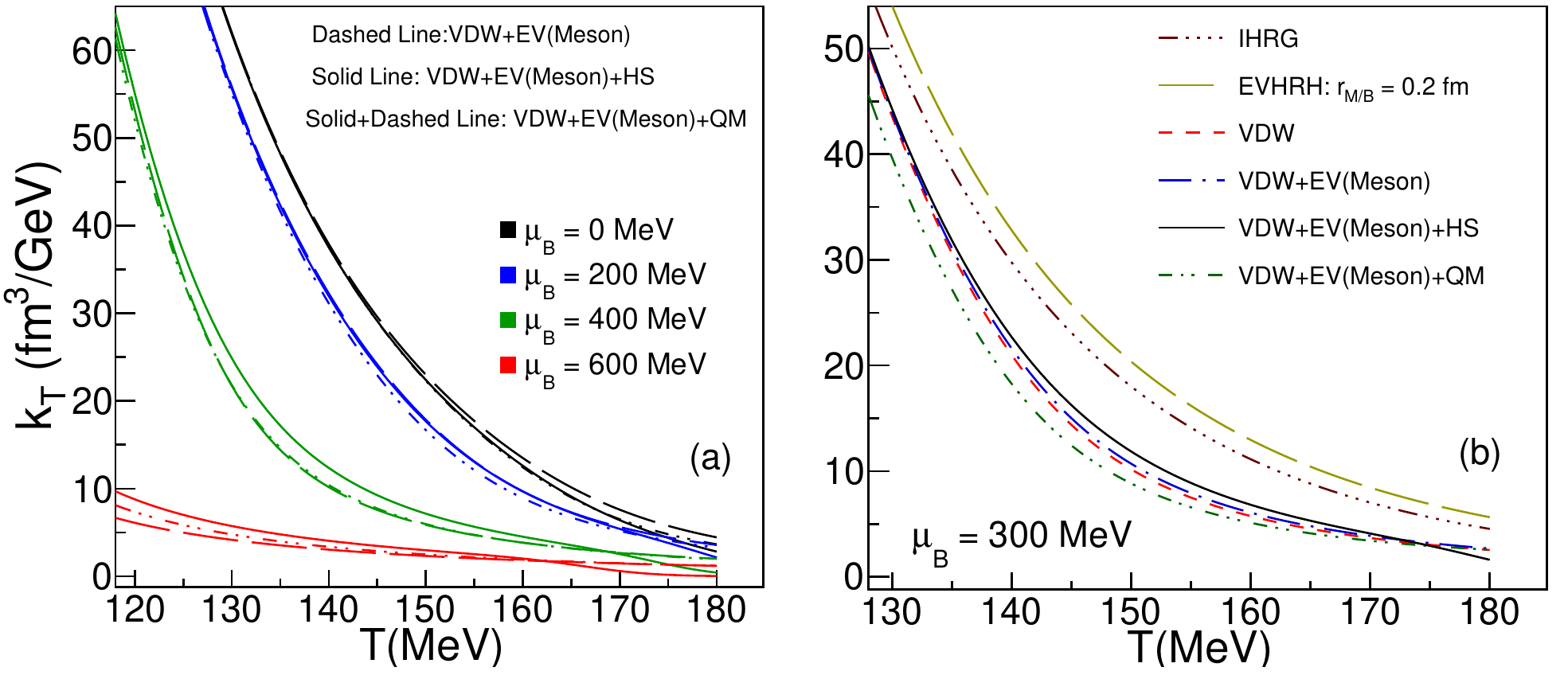} 
	\caption{Left: The isothermal compressibility has been represented as a function of temperature for different VDWHRG models corresponding to four distinct chemical potentials, as specified in the legend. The VDW parameters have been derived from the fitted values of the corresponding VDWHRG model with lattice results at zero and finite chemical potential. Right: Same VDW constants have been used for all the VDWHRG models. The results for IHRG and EVHRG are also provided for comparison.}
	\label{Fig:kTVsT}
\end{figure*}  
Figure \eqref{Fig:kTVsT}-(a) depicts the isothermal compressibility as a function of temperature for different VDWHRG models at four different chemical potentials. The VDW parameters were obtained by fitting the corresponding VDWHRG model at both zero and finite chemical potentials as provided in Tables \eqref{Tab:FiTSumZeroMuB} and \eqref{Tab:FitFiniteMuB}.
As the temperature increases, the number density of the system also increases, causing a decrease in $\kappa_T$. Similarly, as the chemical potential increases, the same trend is observed. The repulsive interaction in the system makes compression more difficult, leading to an increase in $\kappa_T$ \cite{Khuntia:2018non}. At high $\mu_B$ and low temperature, the hierarchy of the $\kappa_T$ values for the different VDWHRG models aligns with the hierarchy of their respective repulsive constants. The interaction effect reduces with smaller chemical potential, and ultimately, the VDWHRG model with more hadronic states (HS or QM) significantly reduces the $\kappa_T$ value at high temperature. Here also, to further clarify the combined effect of the interaction constant and different hadronic states, we plotted $\kappa_T$ for various HRG models with the same van der Waals constant in figure \eqref{Fig:kTVsT}-(b).For completeness, we included IHRG and EVHRG models as well. As expected, QM states have the lowest $\kappa_T$ at low to intermediate temperatures, while HS takes over at high temperatures. The attractive interactions increase the average number fluctuation (see Eq.\eqref{Eq:VDWHRGFlucMeson}), but they also lead to an increase in number density, causing a decrease in $\kappa_T$. This explain why the $\kappa_T$ values for all VDWHRG models are notably lower than those of IHRG or EVHRG. The $\kappa_T$ value for EVHRG is higher than that for IHRG, consistent with previous studies \cite{Khuntia:2018non}. As shown in figure \eqref{Fig:kTVsT}-(b), there is no significant difference in $\kappa_T$ values between VDW and VDW+EV(Meson) models. Therefore, we chose not to include the VDW model calculation for $\kappa_T$ in the figure  \eqref{Fig:kTVsT}-(a) to avoid overcrowding.

\subsection{E. Transport Coefficient: $\eta/s$}

\begin{figure*}
	
	\includegraphics[height=7.0in,width=6.3in]{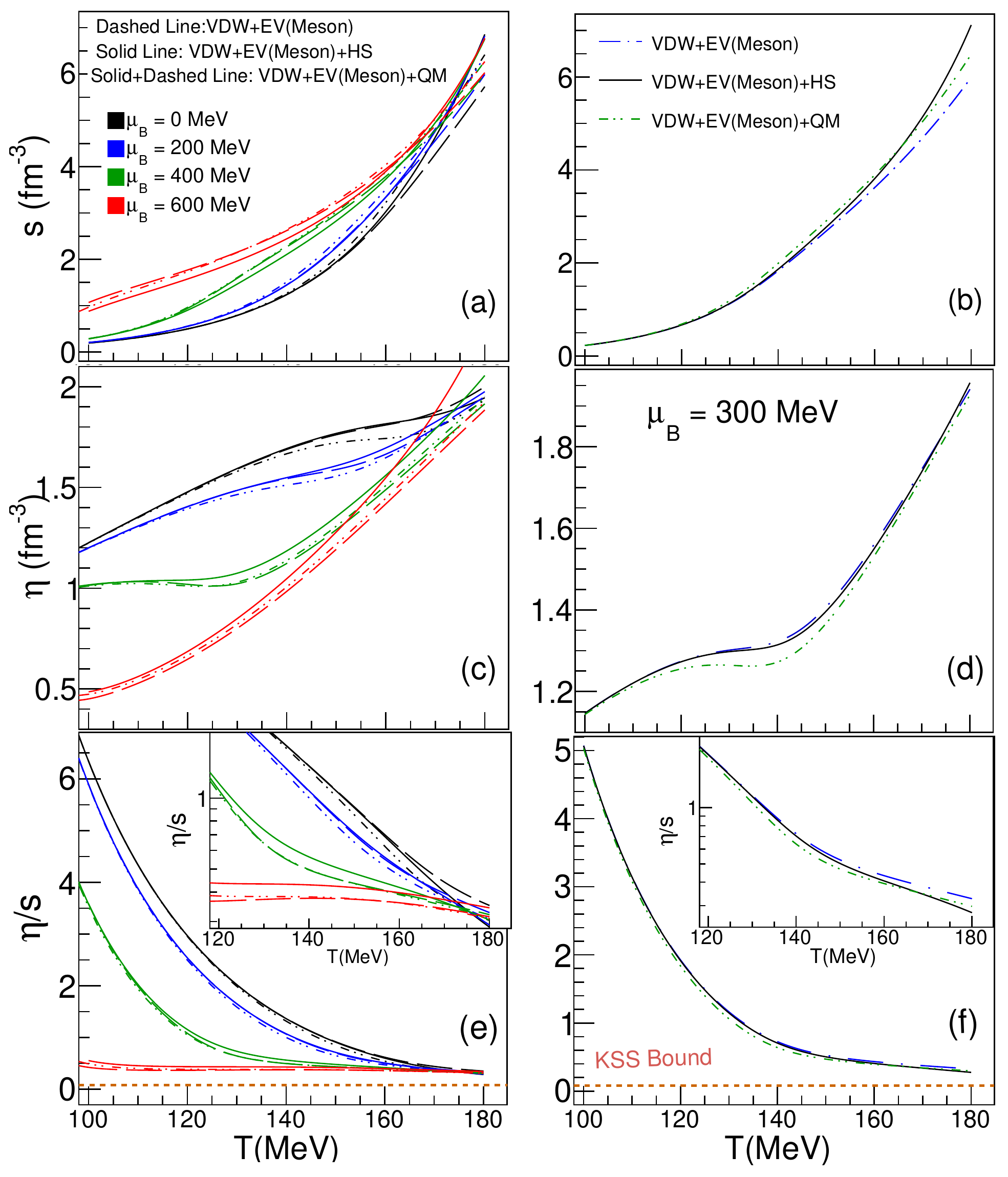} 
	
	\caption{Right Panel: The plot displays the relationship between (a) entropy density, (b) $\eta$, and (c) $\eta/s$ with temperature. Different chemical potentials and variants of the VDWHRG model are represented by various colors and line styles, as noted in the legend. The inset provides a closer view of the high-temperature region, emphasizing the effect of heavier resonance states.
		Left Panel: The plot is similar to the right panel but depicts the relationship between (a) entropy density, (b) $\eta$, and (c) $\eta/s$ with temperature using the same VDW constant for all the HRG models only at $\mu_B$ = 300 MeV. }
	\label{fig:etabys}
\end{figure*}

Next, we investigate the transport coefficient, $\eta/s$, within various VDWHRG frameworks, as examined in this current work. The lower panel of figure \eqref{fig:etabys} illustrates the viscosity to entropy density ratio, $\eta/s$, for different VDWHRG models at various chemical potentials. In order to gain a better understanding of the relationship between $\eta/s$ and its individual components, $\eta$ and $s$, we analyze them separately. The upper panel of figure \eqref{fig:etabys} (first and second row) displays  $\eta$ and $s$, respectively.  Meanwhile, the left panel of the figure \eqref{fig:etabys}-(a,c,e) displays plots of $s$ and $\eta$ as a function of temperature at various chemical potentials for different VDWHRG models, each with the corresponding VDW constants. Conversely, in the right panel, we analyze the same components at a fixed chemical potential of $\mu_B =$ 300 MeV, with the same VDW constant used for all VDWHRG models.

The entropy density and $\eta$ both increase with temperature across all chemical potentials. However, with increasing $\mu_B$, a nontrivial pattern emerges due to the interplay between repulsive and attractive interactions in the high baryon dense region. While at a given low temperature, an increase in number density with decreasing  $\mu_B$ enhances $\eta$ in all considered VDWHRG models. Nonetheless, the entropy density exhibits an opposite behavior, leading to a reduction in $\eta/s$ as $\mu_B$ increases.

At high $\mu_B$ and low temperature, differences in various VDWHRG calculations mainly arise from different VDW constants. Due to the highest repulsive constant for the VDW+EV(Meson)+HS model, the suppression of number density is maximum, resulting in the minimum $\eta$ value. Conversely, the same reason leads to the opposite effect on entropy. As shown in the right panel of figure \eqref{fig:etabys}, using the same VDW parameters for different VDWHRG variants does not result in much difference for both entropy and $\eta$, resulting in a similar $\eta/s$ at low temperatures. However, as expected, due to the presence of extra low mass states, $\eta$ significantly decreases at intermediate temperatures.\\
At low $\mu_B$ and high temperatures, the inclusion of additional states, whether through HS or QM, leads to a reduction in the $\eta$ value. This can be attributed to the fact that the extra states increase the number density, which in turn decreases the viscosity of the system \cite{Pal:2010es,Itakura:2007mx}. Even though the number of counts in the $\eta$ sum increases (see Eq.\eqref{equ:DescritEta}) due to the extra states in the QM model or the Hagedorn states that contribute to the system's viscosity, it is not enough to offset the decrease in $\eta$ caused by the overall increase in the system's number density. Interestingly, at high $\mu_B$  and high temperature, the opposite trend is observed, where the inclusion of extra states leads to a higher shear viscosity.  It can be explained in two aspects.\\
i) At high densities, the strong baryonic repulsion cannot be overcome by including extra states. Therefore, the addition of extra resonances does not have a significant effect on reducing the shear viscosity of the system. \\
ii) At high temperatures, the average momentum, $\Big(\frac{K_{5/2}(m/T)}{K_2(m/T)}\sqrt{\frac{8mT}{  \pi}}\Big)$, of the massive hadrons increases, which leads to a more significant individual contribution of the massive hadrons or HS states to the shear viscosity of the system compared to their contribution to the number density. As expected, the increase of shear viscosity ($\eta$) is maximum for the HS due to the more prominent contribution of continuous states (see Eq.\eqref{equ:HSEta}). However, at both finite and zero chemical potential, the inclusion of additional states leads to an increase in the entropy of the system. Therefore, at high temperatures, the presence of QM-predicted extra hadronic states or the Hagedorn states significantly reduces the value of $\eta/s$ \cite{Noronha-Hostler:2008kkf}. However, at high $\mu_B$, the effect of extra states on $\eta/s$ is minimal, as both $\eta$ and $s$ increase with the inclusion of additional states.

\subsection{F. Collision Energy Dependence }
\begin{figure} 
	\subcaptionbox*{ }[.9\linewidth]{%
		\includegraphics[width=1.01\linewidth]{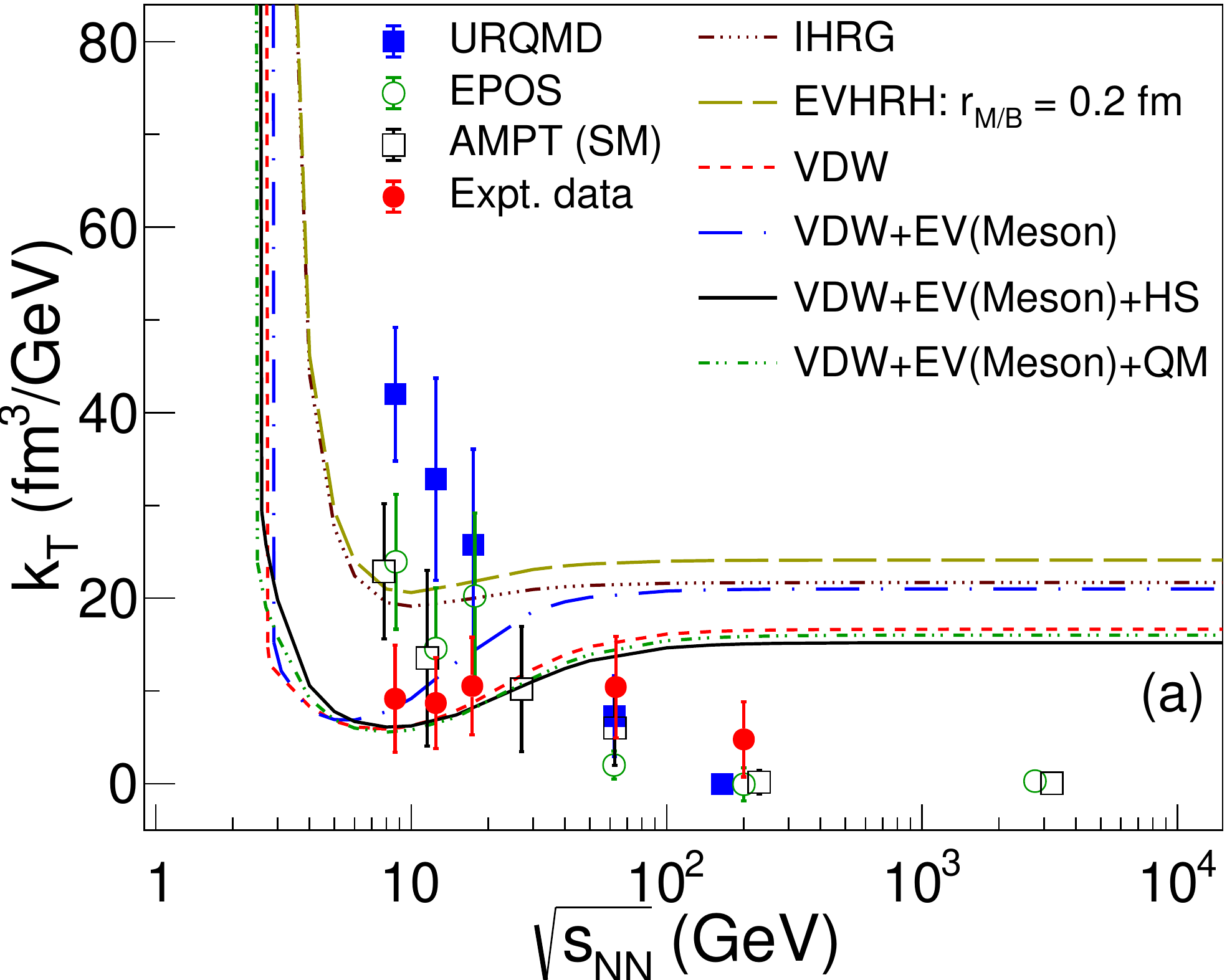}}\\
	\subcaptionbox*{ }[.9\linewidth]{%
		\includegraphics[width=1.01\linewidth]{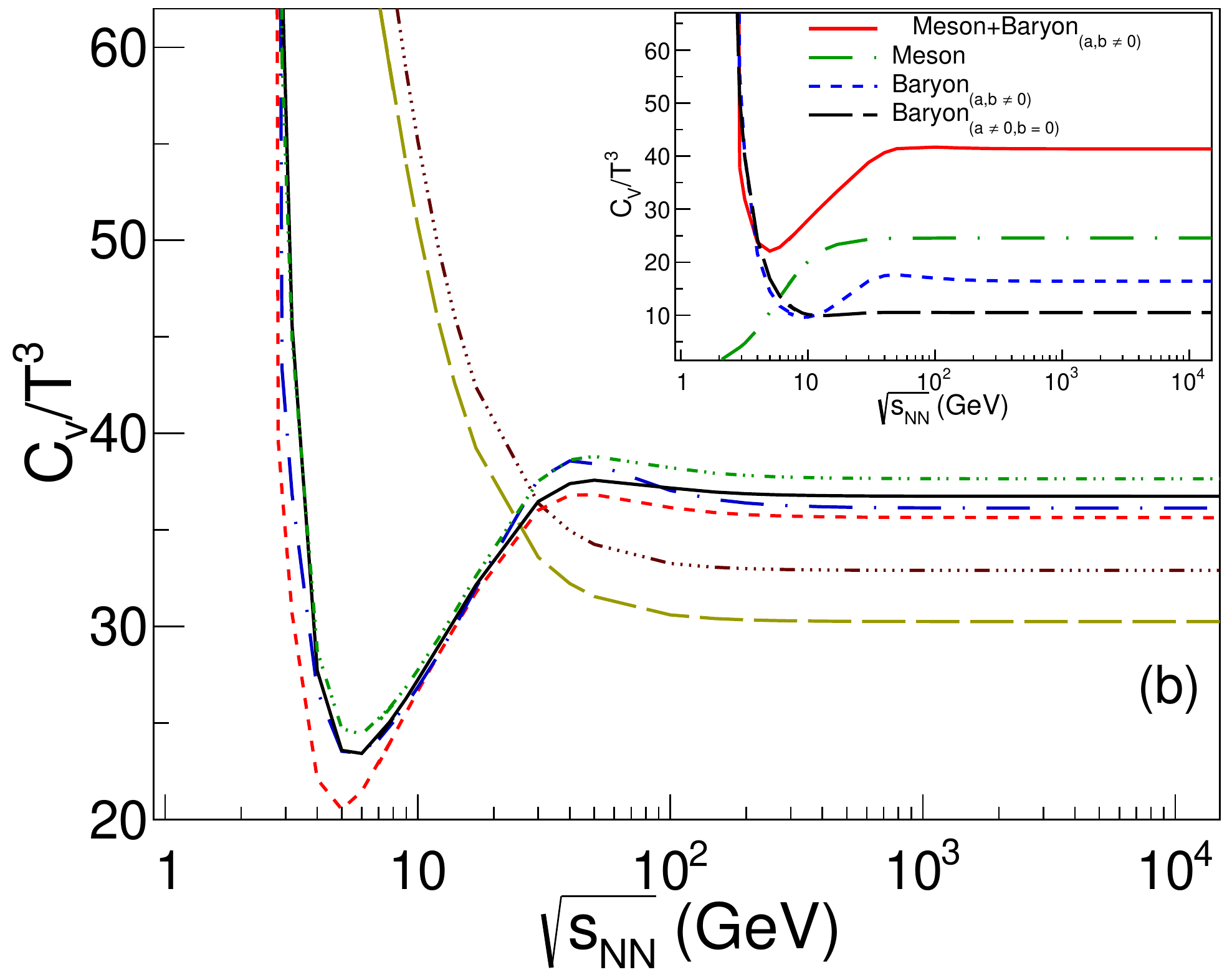}}\\
	\subcaptionbox*{ }[.9\linewidth]{%
		\includegraphics[width=1.01\linewidth]{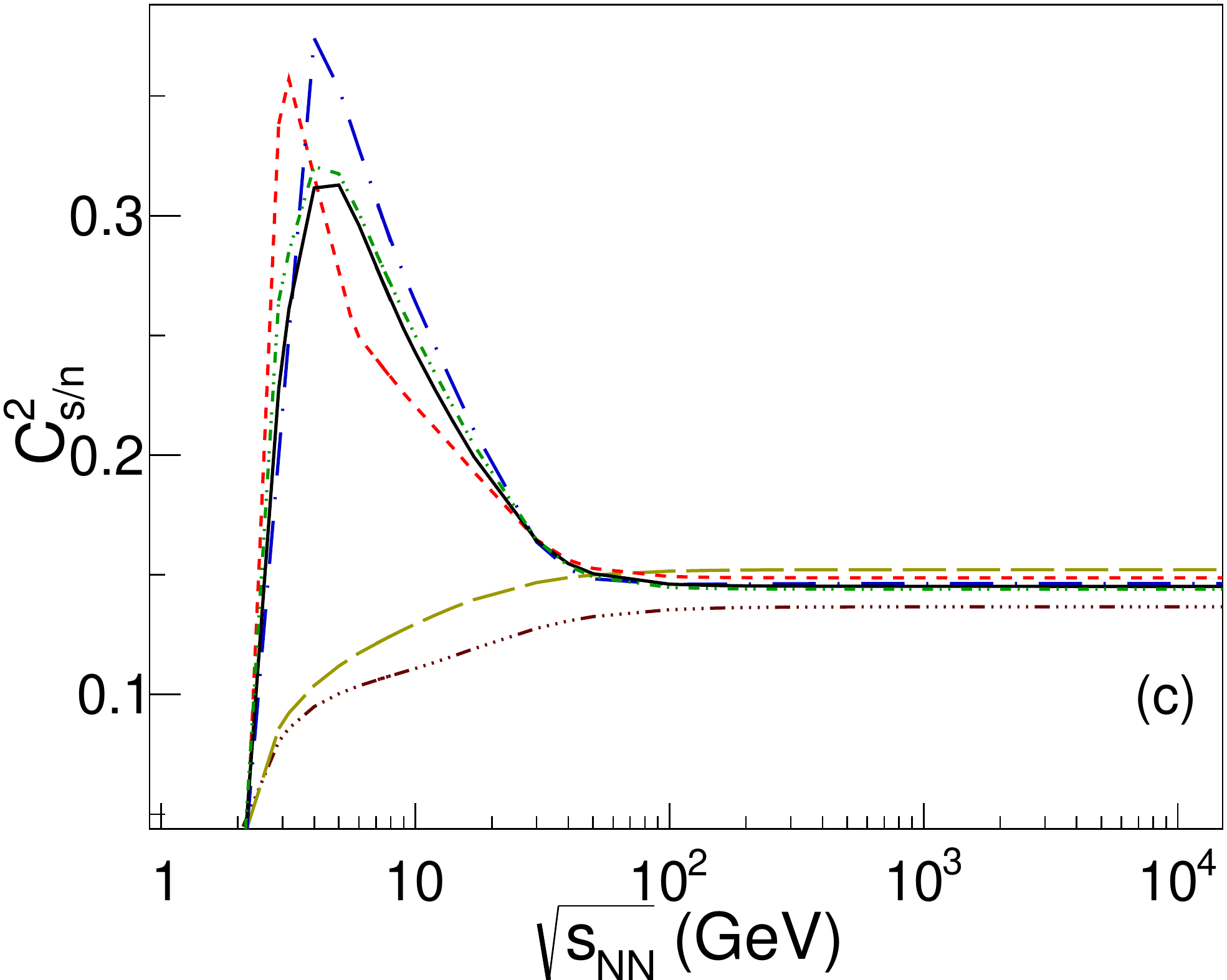}} 
	\caption{ The plot illustrates the dependence of (a) $\kappa_T$ (b) $C_V/T^{3}$, and (c) $C^2_{s/n}$ on $\sqrt{s}$ for various HRG models. (a). Other calculations from references have been superimposed for comparison.
		(b) The inset graph shows the VDWHRG calculation for four different cases, where the details of the variation are explained in the text. The subscripts "a" and "b" represent  the van der Waals parameters.   }
	\label{fig:RootS}
\end{figure}  
In the context of heavy ion collisions, it is important to investigate the dependence of various physical quantities on $\sqrt{s}$. In this study, we examined the $\sqrt{s}$ dependencies of three phenomenologically significant thermal quantities, namely $\kappa_T$, $C_V$, and $C^{2}_s$, at the chemical freeze-out point. To do so, we required freeze-out parameters such as the chemical potential and temperature. Numerous parametric forms of the freeze-out curves have been reported in the literature\cite{Andronic:2005yp,Biswas:2020dsc, Bhattacharyya:2019cer, Andronic:2008gu, Cleymans:2005xv, Poberezhnyuk:2019pxs,Tiwari:2011km}, For our present work, we utilized the following parametric form of the freeze-out curves \cite{Cleymans:2005xv}.

\begin{equation}
	\mu_B =  \frac{a}{1+b\sqrt{s_{NN}}}
\end{equation}

\begin{equation}
	T=c-d\mu^2_B-e\mu^4_B
\end{equation}

The following set of parameter values have been used for the VDWHRG model \cite{Poberezhnyuk:2019pxs}:
$a=1.094\pm 0.004 $ GeV,  $b=0.157\pm 0.002 \  \text{GeV}^{-1}$, $c=0.157\pm 0.002 $ GeV, $d=0.0032\pm 0.0027\  \text{GeV}^{-1}$, and $e=0.259\pm 0.004 \  \text{GeV}^{-3}$.\\
For EVHRG and IHRG we use, 
$a=1.310\pm 0.006 $ GeV,  $b=0.278\pm 0.003 \  \text{GeV}^{-1}$, $c=0.152\pm 0.001 $ GeV, $d=0.026\pm 0.004\  \text{GeV}^{-1}$, and $e=0.219\pm 0.004 \  \text{GeV}^{-3}$.




Figure \ref{fig:RootS}-(a) depicts the center-of-mass energy ($\sqrt{s}$) dependence of the isothermal compressibility ($k_T$) for various HRG models. Consistent with other studies \cite{Mukherjee:2017elm,Khuntia:2018non,Sahoo:2021ljp}, a sharp decline in $k_T$ was observed at low values of $\sqrt{s}$. The EVHRG or IHRG model achieves its minimum value of $k_T$ around $\sqrt{s} = 7$ GeV, with $\kappa_T = 25 \ \text{fm}^3/\text{GeV}$, after which saturation occurs promptly. On the other hand, in all VDWHRG models, the minimum value of $k_T$ is also around the same $\sqrt{s}$, but the corresponding value is notably lower($\sim 7 \ \text{fm}^3/\text{GeV} $), and saturation occurs at a much larger center of mass energy ($\sqrt{s} > 60 \ \text{GeV})$. The presence of attractive interactions in the VDW model may be the reason for this difference.
There were no discernible differences observed among the different VDWHRG models. However, the VDW+EV(Meson)+HS and VDW+EV(Meson)+QM models showed the lowest saturation value because of the additional states they take into account. We also incorporated $k_T$ estimates from other models like URQMD, EPOS, and AMPT-SM (String Melting) in figure \ref{fig:RootS}-(a). It is evident from the figure \ref{fig:RootS}-(a) that the VDW+EV(Meson)+HS model's result is the closest to the $k_T$ estimation obtained from the temperature fluctuation analysis of the experimental data \cite{Mukherjee:2017elm}.

Subsequently, we plotted the scale-specific heat as a function of center-of-mass energy. Here also, we observed a rapid decline in $C_V$ around 30 GeV for the IHRG model and 7 GeV for different VDWHRG models. The values eventually became saturated after $\sqrt{s}>$ 62 GeV, which is in line with the findings reported in \cite{Basu:2016ibk}. The saturated value of $C_{V}/T^3$ in VDWHRG models is higher compared to the IHRG model, due to attractive interactions. An interesting point is a dip in the VDWHRG model calculation at low $\sqrt{s}$, as illustrated in figure \ref{fig:RootS}-(b). To account for this phenomenon, we have inserted a graph of $C_V/T^3$ plotted against $\sqrt{s}$, which examines four distinct scenarios: (i)\ Meson+$\text{Baryon}_{a,b\neq 0}$: involves considering both mesons and baryons, where both repulsive and attractive interactions are present in the baryon sector (ii) Meson: examines only mesons, (iii) $\text{Baryon}_{a,b\neq 0}$: considers only baryons with both repulsive and attractive interactions. Finally, (iv) $\text{Baryon}_{a \neq 0,b = 0}$: only considering baryons with repulsive interactions. It is evident from the inserted figure that the dip occurs due to the combined influence of two factors. Firstly, At that center of mass energy, the system reaches a point where it transitions from a baryon-dominated to a meson-dominated region. Secondly, this point also marks the onset of attractive interactions that start to take effect. Thermal model analysis has already identified this transition from baryon-dominated to meson-dominated freeze-out \cite{Cleymans:2004hj}, which is of significant importance on the characteristics of hadron yields, particularly on the so-called horn structure \cite{Andronic:2008gu}.


Finally, we plotted the variation of $C^2_s$ with $\sqrt{s}$. The IHRG and EVHRG models show a rapid increase in $C^2_s$ that becomes saturated eventually. In contrast, the VDWHRG models exhibit a sharp, inverse dip due to strong baryonic interactions, and the position of the peak differs slightly due to varying VDW constants in the different models. Because of the association of $C^2_s$ with the expansion timescale, maintaining thermal equilibrium becomes very difficult for the system with such high values of $C^2_s$. This may be a signature of the vicinity of the critical point.After $C^2_s$ becomes saturated, no significant difference is observed between the different HRG models.

 \subsection*{G. Location Of The Critical Point}  
 \begin{figure*}[t] 
 	\centering
 	\subcaptionbox*{ }[.48\linewidth][c]{%
 		\includegraphics[width=1.05\linewidth]{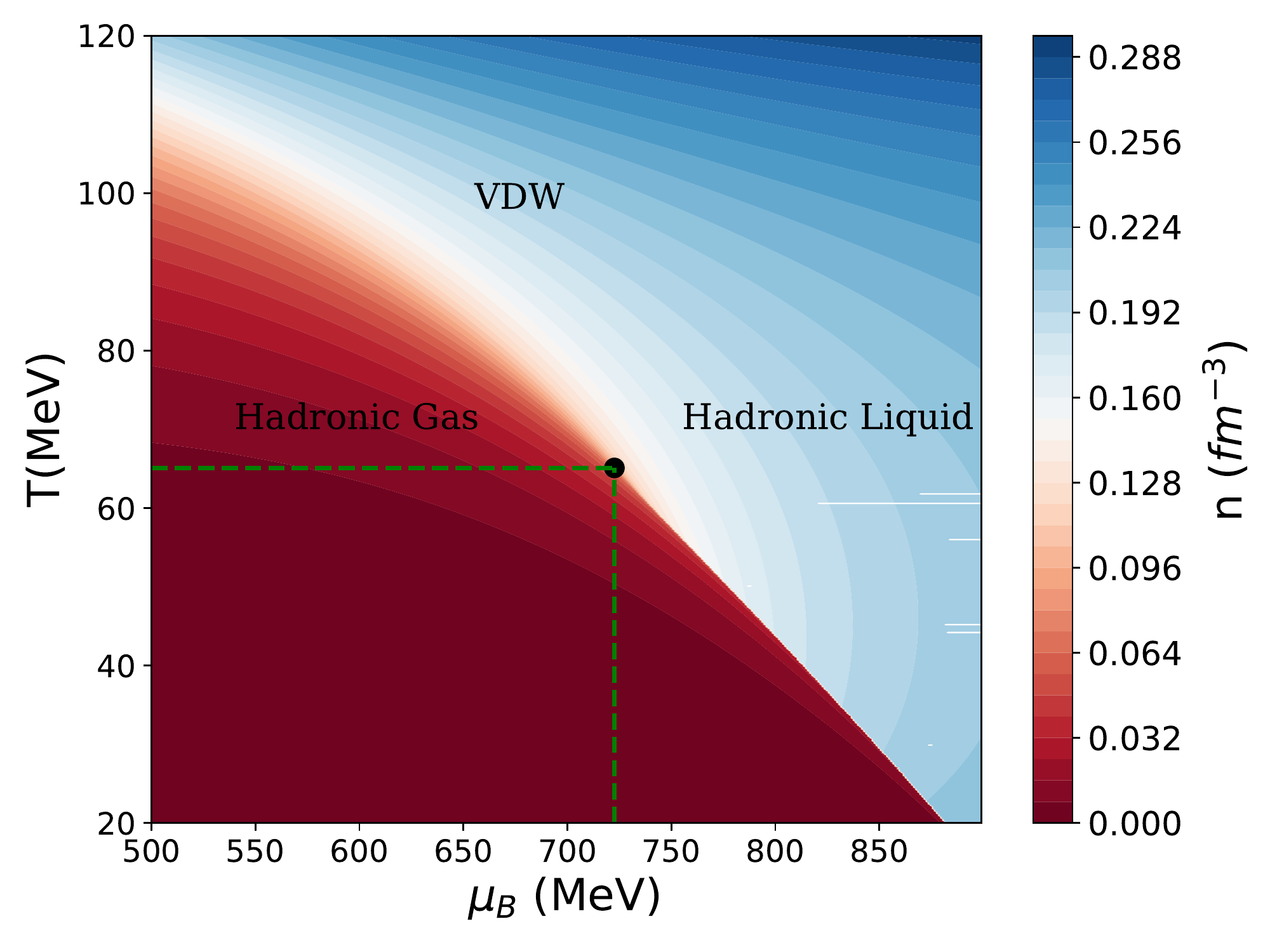}}\quad
 	\subcaptionbox*{ }[.48\linewidth][c]{%
 		\includegraphics[width=1.05\linewidth]{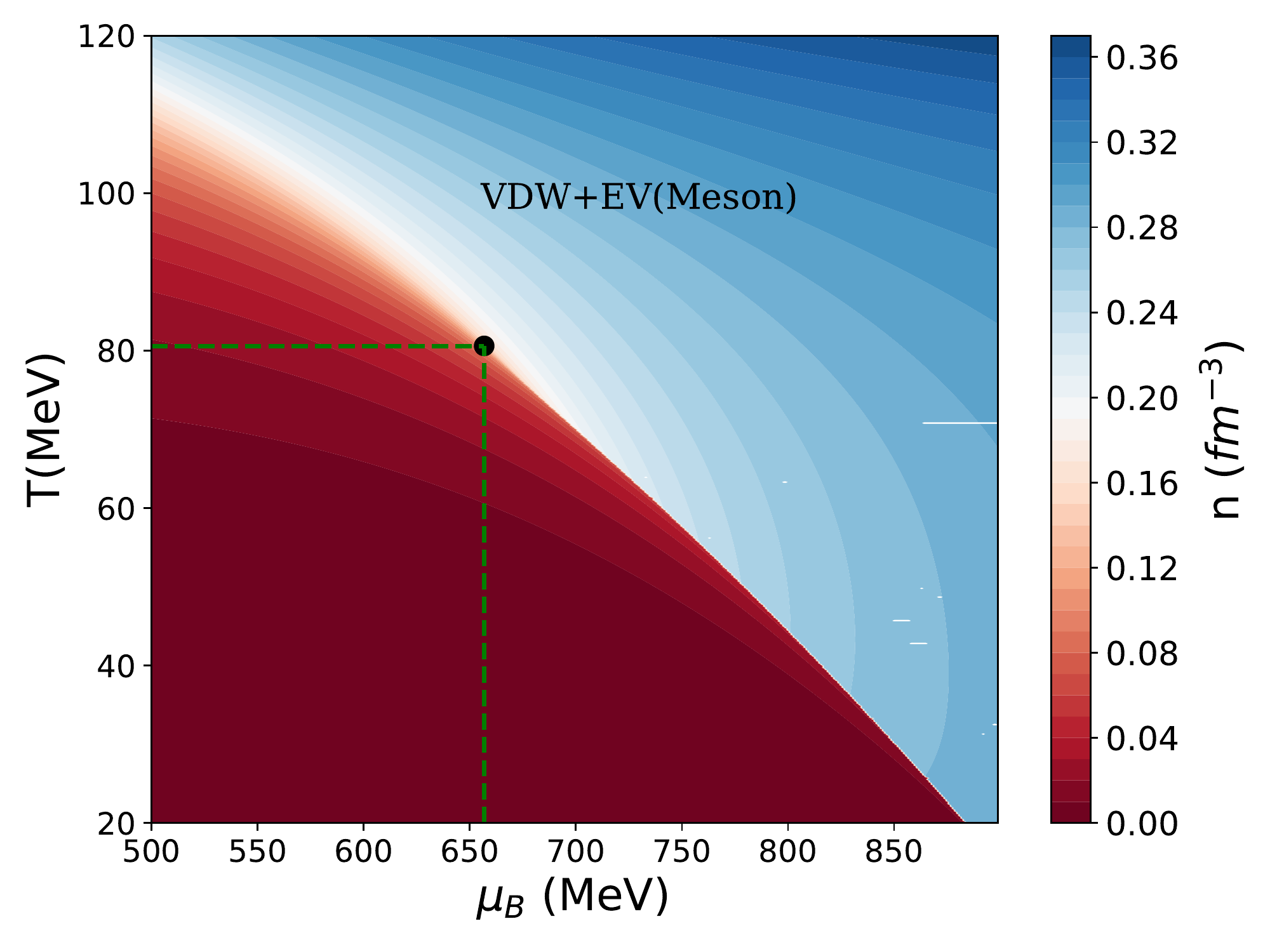}}\quad
 	\subcaptionbox*{ }[.48\linewidth][c]{%
 		\includegraphics[width=1.05\linewidth]{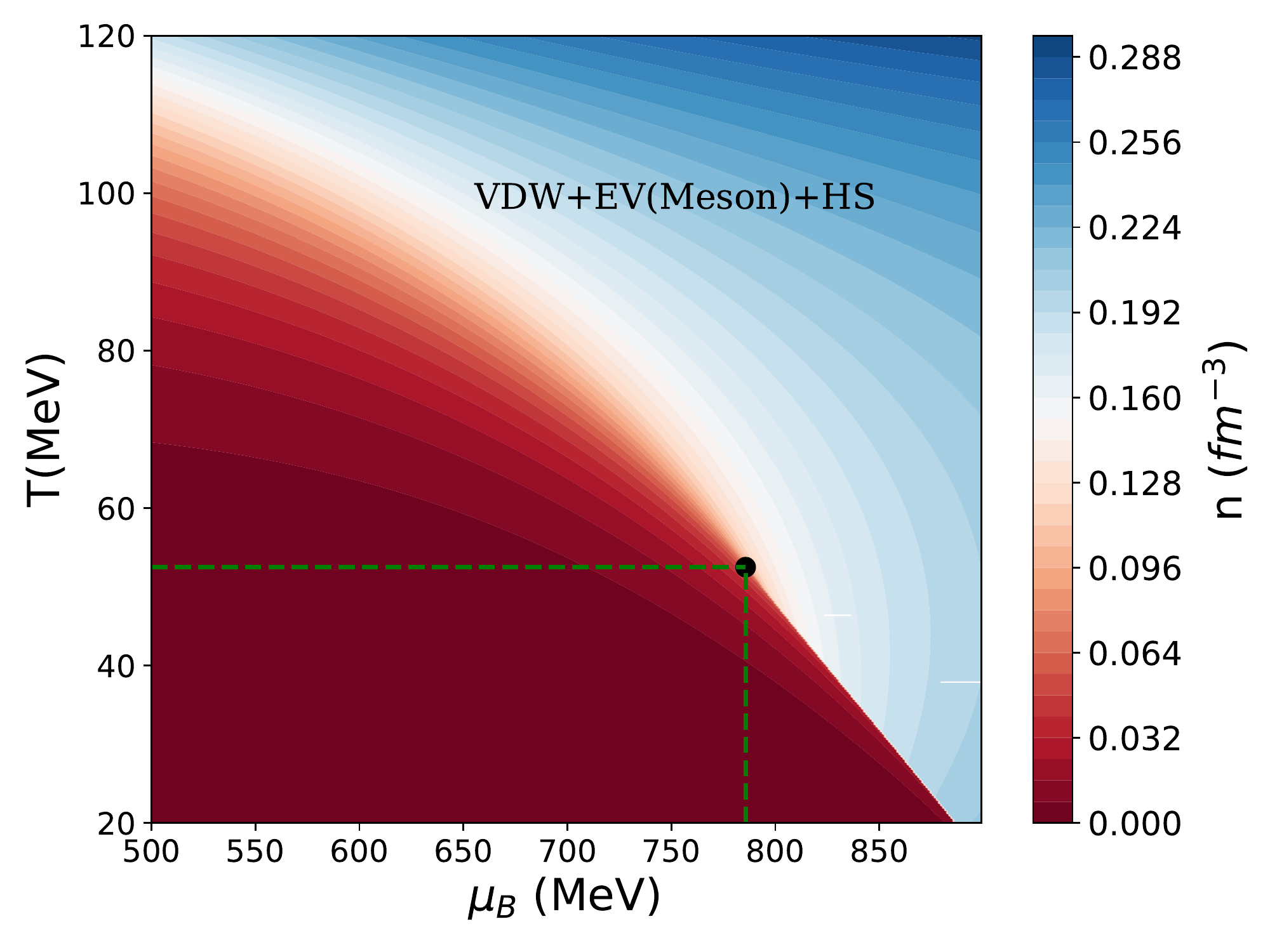}}\quad
 	\subcaptionbox*{ }[.48\linewidth][c]{%
 		\includegraphics[width=1.05\linewidth]{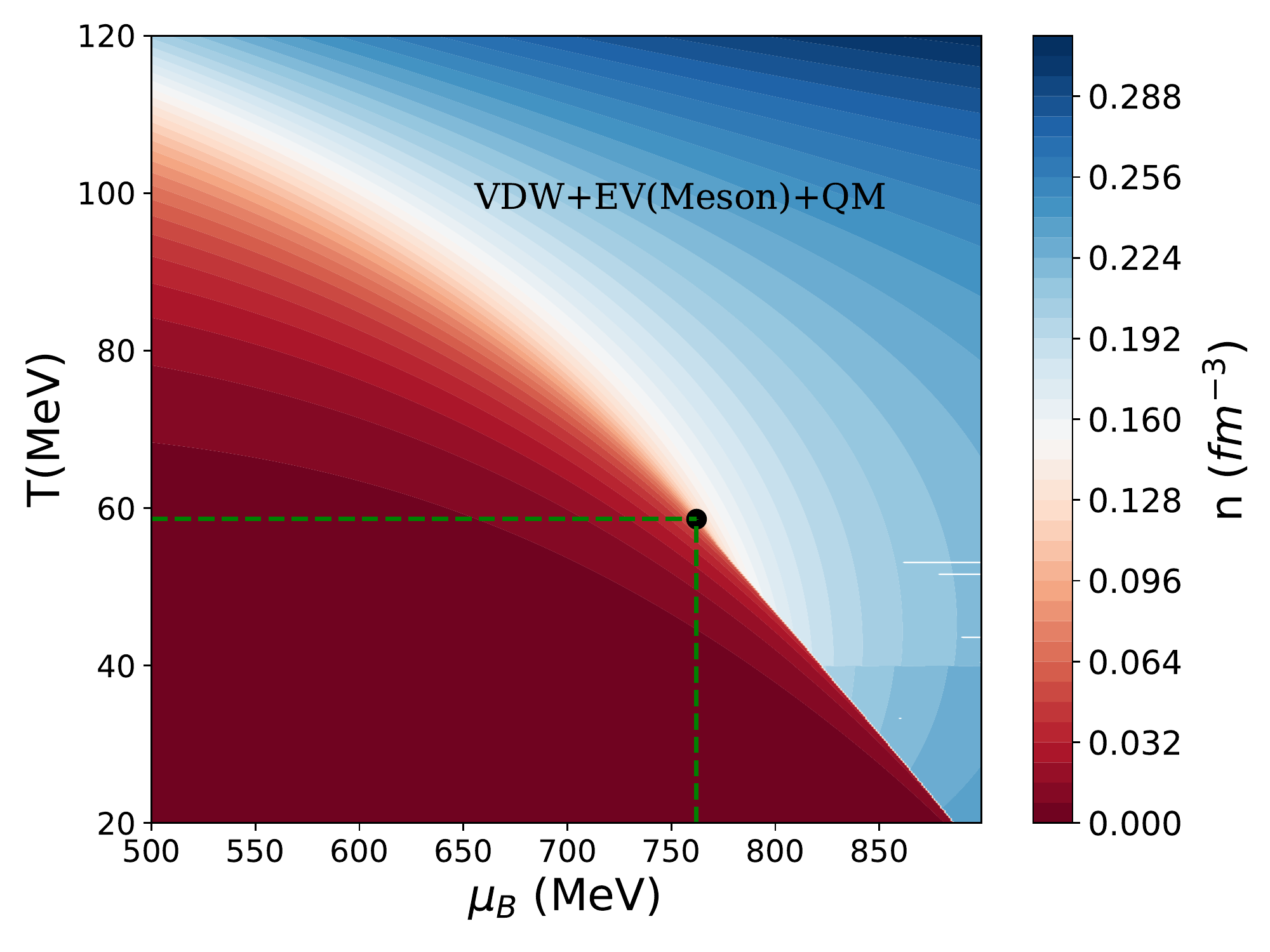}}
 	\caption{ Counter plot of the number density for four distinct VDWHRG models. Black dots indicate the probable location of the corresponding liquid-gas phase transition point.}
 	\label{Fig:CriticalPoint}
 	
 \end{figure*}   
\begin{table}[]
	\centering
	\begin{tabular}[t]{clc}
		\hline

		\hline
		Model & $ \sim \mu^{B}_c$ (MeV)& $\sim T_{c}$ (MeV)   \\
		\hline

		\hline
		
		VDW  & 722.6 &  65.1 \\
		
		\hline
		
		\hline
		VDW+EV &  656.8  & 80.6\\
		
		\hline
		VDW+EV+HS  &  785.7  & 52.5 \\
		\hline
		VDW+EV+QM  & 762.1  & 58.6 \\

		\hline
	\end{tabular}
	
	\caption{Probable location of the critical point for different VDWHRG models.}
	\label{Tab:CriticalPoint}
\end{table}

As previously mentioned, the van der Waals equation incorporates phase transitions and criticality through the selection of van der Waals constants. It is important to note that our study uses the van der Waals interaction, indicating that the observed phase transition is likely a liquid-gas transition, and the obtained critical point corresponds to such a transition. Our objective is to investigate how the choice of VDW parameters in various versions of the VDWHRG model, constrained by LQCD, affects the position of the CP in the QCD phase diagram. To accomplish this, we have generated a contour plot of number density for four distinct cases of the VDWHRG model across the $T-\mu_B$, plane as shown in the figure \eqref{Fig:CriticalPoint} and the probable coordinates of the  critical points for the corresponding models are given in Table \eqref{Tab:CriticalPoint}.It is evident from either the figure \eqref{Fig:CriticalPoint} or the table \eqref{Tab:CriticalPoint} that the locations of the CP  vary significantly depending on the specific VDWHRG model used. These differences mainly arise due to the varying, $a/b$ (Van der Waals parameters) ratios, in the different models. The VDW+EV(Meson) model predicts the highest temperature and lowest $\mu_B$ for the CP, while inclusion of extra states shifts the CP location towards a more baryon-dense region, which could be explored in future experiments involving lower center-of-mass energy collisions.

\section{IV. Summary} 
 
In this study, the we investigated how extra resonance states included through HS or QM affect the van der Waals parameters and their impact on thermodynamic and transport properties of hadronic systems using the VDWHRG model. The van der Waals parameters were tuned through simultaneous fitting of lattice results for various thermodynamic quantities for both zero and finite chemical potential. The results showed that the VDW+EV(Meson) model with Hagedorn states provided the best fit to the lattice results for both finite and zero chemical potential. The inclusion of extra states through HS or QM had a significant impact on the van der Waals parameters, and a strong dependence of the VDW parameters on chemical potential was also observed. The study also found that the repulsive interaction between mesons was important for a good fit at low $\mu_B$, while it was not significant in baryon dense regions.

The temperature profiles of thermodynamic quantities in different VDWHRG models vary due to the combined effects of extra hadronic states and different VDW parameters. VDW parameters have the most prominent influence in the high $\mu_B$ and low temperature range, while extra resonances have more impact with decreasing chemical potential and rising temperature. Specifically, Hagedorn states have the most influence at high temperatures, whereas the inclusion of additional low mass states in the Quark Model has significant effect in the intermediate temperature.

The VDWHRG model exhibits a peak and a minimum in the temperature profile of $C_V/T^3$ and $C_s^2$, respectively, which can be indicative of a QCD phase transition. The location of the peak or minimum is determined by the selected VDW constant. In contrast, no such behavior has been observed in either the EVHRG or IHRG model.
  
As the temperature or chemical potential increases, the isothermal compressibility decreases. In the VDWHRG model, repulsive interactions lead to an increase in $\kappa_T$, while attractive interactions lead to a decrease. The presence of attractive interactions in the VDWHRG model results in significantly lower $\kappa_T$ values compared to the IHRG or EVHRG models.
 
The transport coefficient, $\eta/s$, is influenced by both the van der Waals (VDW) constant and the presence of extra resonance states as well. Most notably, at low values of baryon chemical potential and high temperatures, the inclusion of extra resonance states causes a significant decrease in the $\eta/s$ ratio. Conversely, at high $\mu_B$, the strong repulsive effect of the VDW constant counteracts the impact of the additional states on $\eta/s$.

We finally linked our findings to heavy-ion collision experiments by studying the  $\sqrt{s}$ dependence of thermodynamic observables. All HRG models showed a substantial drop in $k_T$ at low $\sqrt{s}$, which stabilized at higher $\sqrt{s}$. However, the relationship between $k_T$ and $\sqrt{s}$ differed significantly in VDWHRG models compared to EVHRG or IHRG models. Among the VDWHRG models, the VDW+EV(Meson)+HS model provided the closest estimate of $k_T$ to the value obtained from temperature fluctuation analysis of the experimental data. The dip in $C_V/T^3$ observed at a specific collision energy can be explained by the combined effect of the system transitioning from a baryon-dominated to a meson-dominated region and the onset of attractive interactions at that point. Conversely, an inverse dip was observed in $C^2_s$.

We have also observed that the probable location of the critical point varies significantly among the different VDWHRG models considered here. Models with extra states predict the location of the CP to be in a more baryon dense region, corresponding to very low $\sqrt{s}$.

In Conclusion, the inclusion of extra resonance states through HS or QM has a significant impact on the VDW parameters and, consequently, on the thermodynamic and transport properties of the hadronic system, Moreover, this impact varies over a broad $T-\mu_B$ range in the QCD phase diagram. Therefore, it is crucial to properly tune the VDW constant, taking into account these extra states (HS or QM), before proceeding with further analyses, such as yield or fluctuation analyses, within the VDWHRG framework.

\section{Acknowledgments} 
 I am thankful to Deeptak Biswas, Amaresh Jaiswal, and Unmesh Dutta Chowdhury for their meticulous review of the manuscript and their valuable suggestions. Additionally, I would like to extend my thanks to Deeptak Biswas and Hiranmaya Mishra for their useful discussions.
 
 \bibliographystyle{unsrt}
 \bibliography{VDWHRG_With_HS_QM.bib}

\end{document}